\documentclass[12pt,preprint]{aastex}
\newcommand{\be}{\begin{equation}}
\newcommand{\ee}{\end{equation}}
\newcommand{\sssty}{\scriptscriptstyle}
\newcommand{\lb}[1]{\label{#1}}
\newcommand{\apg}{\:^{>}_{\sim}\:}
\newcommand{\apl}{\:^{<}_{\sim}\:}
\newcommand{\etal}{{\it et al.\ }}
\shorttitle{Number Counts and Luminosity Function}
\shortauthors{Ribeiro and Stoeger}
\begin{document}
\title{Relativistic Cosmology Number Counts \\ and the Luminosity
       Function}
\author{Marcelo B.\ Ribeiro\altaffilmark{1} and William R. Stoeger}
\affil{Vatican Observatory Group, Steward Observatory, \\ University of
       Arizona, Tucson, AZ 85721, USA}
\email{mribeiro@as.arizona.edu, wstoeger@as.arizona.edu}
\altaffiltext{1}{On leave from Physics Institute, University of
      Brazil - UFRJ, Rio de Janeiro.}

\begin{abstract}
This paper aims to connect the theory of relativistic
cosmology number counts with the astronomical data, practice, and
theory behind the galaxy luminosity function (LF). We treat galaxies
as the building blocks of the Universe, but ignore most aspects of
their internal structures by considering them as point sources.
However, we do consider general morphological types in order to use
data from galaxy redshift surveys, where some kind of morphological
classification is adopted. We start with a general relativistic
treatment for a general spacetime, not just for
Friedmann-Lema\^{\i}tre-Robertson-Walker, of number counts, and
then link the derived expressions
with the LF definition adopted in observational cosmology. Then
equations for differential number counts, the related relativistic
density per source, and observed and total relativistic energy
densities of the universe, and other related quantities are written
in terms of the luminosity and selection functions. As an example of
how these theoretical/observational relationships can be used, we
apply them to test the LF parameters determined from the CNOC2 galaxy
redshift survey, for consistency with the Einstein-de Sitter (EdS)
cosmology, which they assume, for intermediate redshifts. We conclude
that there is a general consistency for the tests we carried out, namely
both the observed relativistic mass-energy density, and the observed
relativistic mass-energy density per source, which is equivalent to
differential number counts, in an EdS Universe. In addition, we find
clear evidence of a large amount of hidden mass, as has been obvious
from many earlier investigations. At the same time, we find that the
CNOC2 LF give differential galaxy counts somewhat above the EdS
predictions, indicating that this survey observes more galaxies
at $0.1 \apl z \apl 0.4$ than the model's predictions.
\end{abstract}


\section{Introduction}\lb{intro}

In cosmology, it is one thing to obtain universe models from
solutions of Einstein's field equations, and quite another to express
these solutions in terms of quantities and parameters which can be
related directly to real astronomical observations. This intrinsic difficulty
of cosmological research was perceived long ago by its practitioners,
more or less by the time modern cosmology was launching its own
foundations, in the 1920's and early 1930's. This perception initiated
a process which led to the establishment of a division between
theoreticians and observers, with the former being grouped in what may
be termed as {\it relativistic cosmology}, usually considered to be the 
realm of the first task above, and the latter forming {\it
observational cosmology}, which lies in the realm of the second one.
Despite this division, there has always been a few,  of course,
who were able to link successfully these two sub-fields ({\it e.g.},
Tolman 1934; McCrea 1934, 1939; Sandage 1961, 1995; Ellis 1971;
Weinberg 1972; Ellis and Perry 1979; Ellis {\it et al.\ }1984;
Peebles 1980, 1993; Longair 1995). 
And more recently, there have been many individuals and
groups of researchers who have successfully and closely linked cosmological
observations (particularly those of the cosmic microwave background
radiation and those involving deep galaxy surveys) to the standard
Friedmann-Lema\^{\i}tre-Robertson-Walker (FLRW) theory. However, more
needs to be done, particularly with regard to the broader issues of
using observations to determine how good a model FLRW is on various
cosmological length scales.

Each of these two approaches to cosmology (relativistic-theoretical
and observational) has its own distinctive features.
Relativistic cosmologists have been so successful in coping with
the difficulties of obtaining exact solutions of Einstein's field
equations, that today one of the problems of this area is not the
lack, but rather the excess, of solutions without the correspondent
astrophysical modeling (Krasi\'{n}ski 1997). So, in relativistic
cosmology a different cosmological model means in fact a different
spacetime metric. Observational cosmologists, on the other hand,
usually work almost exclusively with FLRW spacetimes, since, by
its very nature, observational verification in cosmology is a
formidable task, and since there are indications that the Universe is nearly
FLRW on the largest scales. The fact that the characterization of FLRW geometry
can be reduced to a set of a few parameters to be determined
observationally, has allowed observational verification in cosmology
to be reduced to manageable proportions. Or so it has been thought.
Consequently, for observational cosmology a different cosmology often means,
in fact, different parameter values for an FLRW spacetime
metric, changing, for instance, its general topology from flat, to closed or
open. Nevertheless, however different, these two sub-areas share the
same general purpose of modern cosmology, which is to determine the
spacetime geometry and the mass distribution of the Universe.

One of the most recent and comprehensive attempts to unify these two
approaches in cosmology has been,
perhaps, the {\it ideal observational cosmology} program, which
proposes to characterize in detail the way in which cosmological
observations can be directly used to determine the cosmological spacetime
geometry (Ellis{ \it et al.\ }1985; Nel 1987). The main motivation
behind this program has been to determine the spacetime metric of our
universe directly from astronomical observations, without assuming a
cosmological model beforehand (see Ellis, Matravers and Stoeger 1995).
In the process of doing that, one must first determine what is and
what is not decidable in cosmology on the basis of astronomical
observations (Ellis{ \it et al.\ }1985, p.\ 317).

However, despite its obvious appeal, the ideal observational cosmology
program still encounters a very basic problem: the astrophysical
evolution, both in number and in luminosity, of galaxies, which are
the fundamental astronomical makers of cosmic spacetime. Astrophysical
evolution and cosmology are convolved in every presently practical
astronomical observation of cosmological relevance. In order to
implement the ideal observational cosmology program, not only do we
require precise observations of redshift, galaxy number counts, and
observer area distances (or luminosity distances), in the spherically
symmetric case, but we also need an adequate model of how the populations
and luminosities of the different types of galaxies evolve with redshift.
It is presently impossible to determine this without assuming some
cosmological model. This problem is so difficult that astronomers
usually assume that the universe is well described by the standard
FLRW geometry, and analyze, interpret and present their observational
results in terms of that class of models, with or without a cosmological
constant. On the largest scales -- that is, on the largest scales probed
by the cosmic microwave background radiation (CMBR) -- there is good
reason for doing this (see Stoeger, Maartens and Ellis 1995), although
on scales less than these, this assumption is still in need of more
adequate justification, inasmuch as we are incapable so far of
determining the smallest length scale on which the FLRW approximation
is valid. Thus, most observational results in cosmology
either explicitly or implicitly assume an FLRW model in their presentation.

Although this approach has enabled the rapid progress of cosmology to this
point, it may significantly hamper taking the next step: determining
observationally how significant the deviations from FLRW are on various
length scales, once we have some adequate model of galaxy evolution.

Finally, in some cases, in order to obtain simplified models, the
approach taken in observational cosmology -- particularly at low
redshifts -- has not been fully relativistic, and sometimes involves
Euclidean simplifications, without proper verification of the limits
within which such simplifications are justifiable (Ellis 1987; Ribeiro
2001). Though with the advent of much deeper surveys this is very rare
now, it still causes some confusion, since some standard references
show simplified presentations which, often for clarity, do not include
these corrections.\footnote{ \ A common Euclidean approximation used
in the past in cosmology is to calculate the galaxy number counts by
means of an energy flux density which is not corrected by the redshift
(see, {\it e.g.}, Peebles 1993, p.\ 214). Another commonly used Euclidean
approximation was to take the mass density $\rho$ to be constant when
$z \le 0.1$ without a proper verification if that is theoretically
 advisable within the problem under discussion. Although such
approximations, often used in the past, are a rarity in recent redshift
surveys analysis, the fact that they can be found in often-quoted texts
is still a source of confusion.}

The aim of this paper is to bridge the relativistic and observational
cosmology approaches. But, it is much less ambitious than the ideal
observational cosmology program, although we do intend to start with a
fully relativistic approach and go as far as possible with the general
equations, only then specializing them to particular geometries. In
this paper, as an important example, we specialize to the Einstein-de
Sitter model. The basic philosophy follows the general proposal for
limited bandwidth observations of cosmological point sources outlined
in Ribeiro (2002). However, our focus here will be on galaxy number
counts, since this a very important cosmologically relevant
quantity, giving information about the density of mass-energy in the
universe, and can be determined by careful analysis of the data in
galaxy redshift surveys. More particularly, therefore, our purpose is
to present the detailed relationship between galaxy number counts, or
galaxy luminosity functions, and relativistic mass-energy density and
relativistic mass-energy density per source of the universe for any
general cosmological model, and, as an example of how these relationships
can be used, specialize them to FLRW and employ them to test the
consistency of the FLRW model, in this case Einstein-de Sitter (EdS),
assumed in deriving luminosity functions from a given redshift survey
(the CNOC2 intermediate redshift survey) with the mass-energy density
and mass-energy density per source, as functions of redshift, implied
by that model. Do the number counts really determine these parameters
to be EdS, as they should?

There are other ways, of course, in which we can use this framework.
For instance, investigating the interplay of
evolution, selection-effect and incompleteness models with various
cosmologies in interpreting the same data, and testing simplifying
and non-relativistic assumptions for certain redshift regimes. It should
be mentioned here, too, that the other important cosmologically relevant
observational parameter which complements number counts in determining
the cosmological and evolution models is observer area distance (or
luminosity distance). An scheme analogous to the one we present here
for number counts can be established for this parameter (see Stoeger
1987; Stoeger and Ara\'ujo 1999).

Due to the reasons stated above, we shall in this paper specialize
our model to FLRW. But we intend to expand this study eventually to
non-FLRW metrics. This is justified because we wish to see what
are the main differences between the predictions of reasonable non-FLRW
metrics and those of the standard (FLRW) model. The only way we can
really determine how good the Friedmann models are, is to fit almost-FLRW
and non-FLRW models to the data, and compare them with the Friedmann
fits to see if these latter are better (Ribeiro 2002).
As the next logical step the obvious choice would be the
Lema\^{\i}tre-Tolman-Bondi spacetime, since this is the most widely
used metric for testing cosmological models after the standard model
(Krasi\'{n}ski 1997).

Much of this detailed theoretical relativistic treatment has been
presented before, but it was never linked or applied directly to
luminosity function (LF) and number count data, with a careful discussion
of the problems and obstacles of doing so. This is the first in a series
of papers investigating how a fully rigorous general theoretical
cosmological framework can effectively incorporate cosmologically
relevant observational data -- first assuming  FLRW and testing for
consistency, and then testing the FLRW hypothesis itself on various
cosmologically relevant length scales by seeing if non-FLRW models
also fit the data, with reasonable evolution, selection-effect and
incompleteness assumptions.

As mentioned above, since most cosmological data is analyzed and
presented employing the assumption of FLRW, we shall first perform
consistency tests of the derived LF's with the FLRW model they assume.
Models with non-zero cosmological constant will be discussed in a
follow-up paper (Stoeger and Ribeiro 2003)

The main conclusions we reach in this study is that the luminosity
function parameters derived from data acquired by the recent redshift
surveys are roughly consistent with the theoretical predictions of the
standard Einstein-de Sitter (EdS) model. However, the relativistic
energy density per source, or equivalently the galaxy number counts,
departs from this model's predictions at $0.1 \apl z \apl 0.4$. That
may indicate either an inadequacy of the EdS cosmology to represent
the data at those redshifts, or the rough number evolution model used
in the CNOC2 LF calculations needs improvements.

This paper is organized as follows. In \S \ref{model} we derive metric
independent relations based on number source counting and the luminosity
function. In \S \ref{luminosity} we collect the available data on the
luminosity function and organize them in a form usable to our purposes.
Section \ref{flrw} specializes the general equations to the Einstein-de
Sitter spacetime geometry, whereas section \ref{testing} carries out two
consistency tests between the observational data and Einstein-de
Sitter's cosmological model theoretical predictions. Section
\ref{concluding} discusses the results and their implications.

\section{Model Independent Counts of Cosmological Sources}\lb{model}

We begin by deriving completely general cosmological
number counting dependent quantities, which could, in principle, be
determined directly from observations.

\subsection{Relativistic Density of the Universe}

\subsubsection{Relativistic Number Counts}

\begin{figure}
\plotone{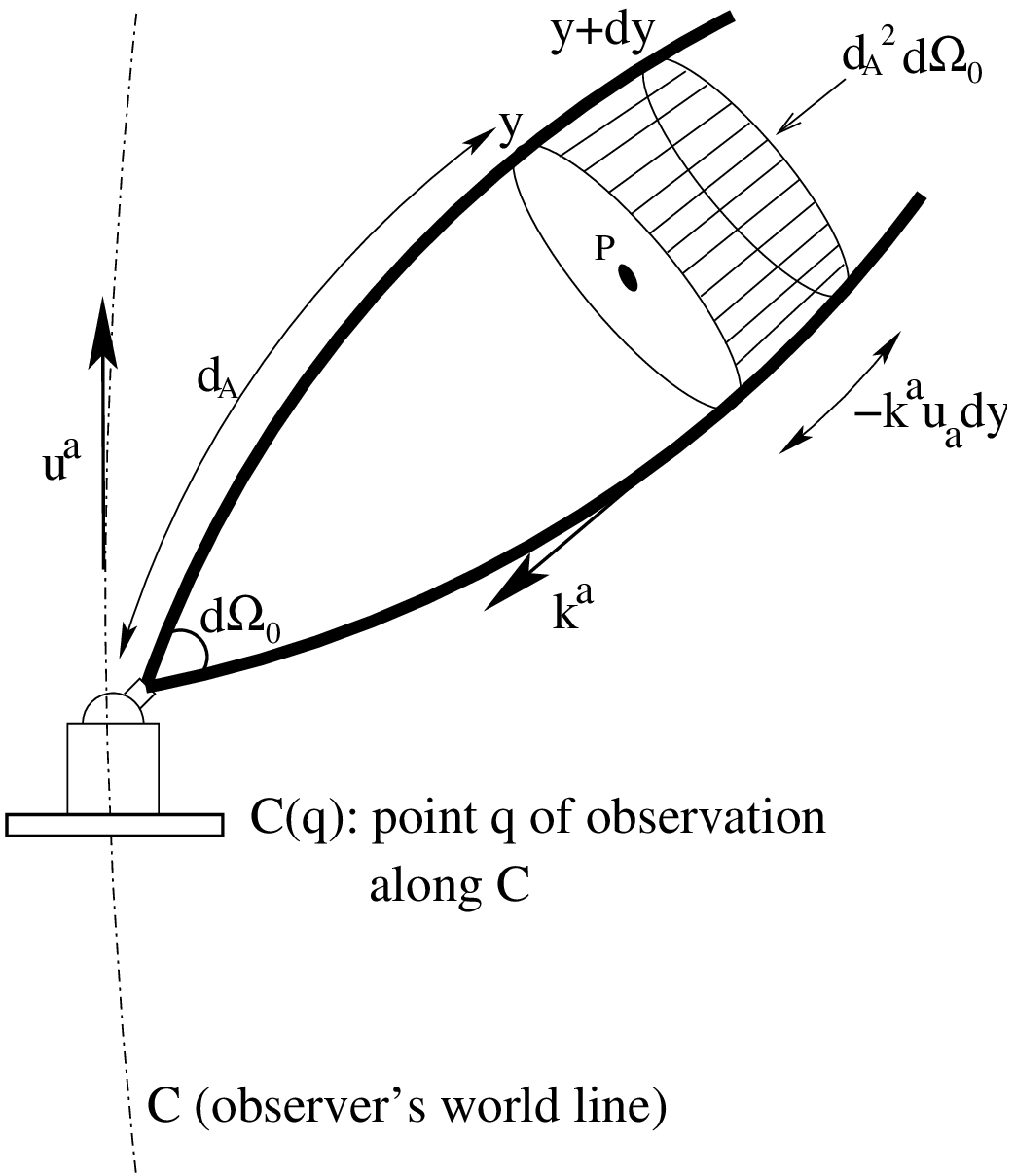}
\caption{Section of a bundle of light rays subtending a solid angle
        $d \Omega_0$ as seen by the observer in $C(q)$. The affine
	parameter displacement $dy$ corresponds to a local distance
        variation of $\left( -k^au_a \right) dy$ in the
        rest-frame of the galaxy at a point $P$ down the null cone
       (Ellis 1971).} \label{fig1}
\end{figure}

The key result derived by Ellis (1971) for the {\it 
number of cosmological sources} in a volume section at a point $P$ down
the null cone is
\be dN= ({d_{\scriptscriptstyle A}})^2 d \Omega_0 { \left[ n
    \left( -k^a u_a \right) \right] }_P \; dy. 
    \lb{dn}
\ee
Here $n$ is the {\it number density} of radiating sources per unit
proper volume in a section of a bundle of light rays converging towards
the observer and subtending a {\it solid angle} $d \Omega_0$ at the
observer's position, $d_{\scriptscriptstyle A}$ is the {\it area
distance}~\footnote{ \ In the literature $d_{\scriptscriptstyle A}$ is
also known as {\it angular diameter distance}, {\it observer area
distance}, and {\it corrected luminosity distance} (Ribeiro 2001,
footnote 9, p.\ 1711).} of this section from the observer's viewpoint,
$u^a$ is the {\it observer's 4-velocity}, $k^a$ is the {\it tangent vector
along the light rays}, and $y$ is the {\it affine parameter distance}
down the light rays constituting the bundle (see figure \ref{fig1}).
Notice that $n$ is measured in the rest-frame of the counted sources.
This equation is metric independent, being, therefore, valid for
{\it all} cosmological models (see Ellis 1971, p.\ 159 for more details).
It should be noted that, in general, $n$ has a dependence both on the
cosmology and on the number evolution of the sources. Only when the
density of galaxies is constant in a comoving volume will $n$ be free
of number evolution.

While the equation above is precise, it is written in terms of
unobservable quantities, like the affine parameter $y$. It can be
expressed in terms of observable quantities, but this has been
previously done in observational coordinates, which are quite
different from the usual ones (Ellis{ \it et al.\ }1985).

The general definition for the {\it redshift} can be written as
(Ellis 1971, p.\ 146),
\be
 1+z=\frac{{ \left[ u^a k_a \right]}_{\mbox{\footnotesize source}}}
          {{ \left[ u^a k_a \right]}_{\mbox{\footnotesize observer}}}
    =\frac{{ \left[ u^a k_a \right]}_P}
          {{ \left[ u^a k_a \right]}_{C(q)}}.
 \lb{z}
\ee
Substituting equation (\ref{z}) into equation (\ref{dn}) yields,
\be
  dN= ({d_{\scriptscriptstyle A}})^2 d \Omega_0 { [n]}_P (1+z) { \left[
      \left( -k^a u_a \right) \right] }_{c(q)} \; dy.
  \lb{dn2}
\ee

All the light rays reaching the observer's worldline $C$ at the same time
form a hypersurface which we call a {\it past light cone}, and which can
be represented by a constant quantity over this surface. Following the
notation of Ellis' \etal (1985, p.\ 324), let us call the function
generating this surface $w$. Therefore,
\be w=w(x^a)=\mbox{const}, \lb{w} \ee
defines the past light cones along the observer's worldline, the locus
of light rays reaching the observer at a given moment of time. This
constant is arbitrary, and we will choose it to be equal to the
observer's proper time $\tau$ at its worldline $C$. So, we may write the
following equation,
\be { \left[ w(x^a) \right] }_C = { \left[ \tau \right] }_C. \lb{w2} \ee
If we now define a vector field $k_a$, such that,
\be k_a \equiv - \frac{d w}{dx^a}, \lb{k} \ee
then this vector field generates past light cone surfaces. The minus sign
was chosen because we are interested in {\it incoming} light rays, which
are indeed geodesics of null type if their generating vector field obeys
the following conditions (Ellis 1971; Ellis{ \it et al.\ }1985;
Schneider{ \it et al.\ }1992, p.\ 94),
\be k^ak_a=0, \; \; \; {k^a}_{;b} k^b=0. \lb{k2} \ee
Now, using the definitions of $k_a$ and $u^a$, it follows that, 
\be k_a u^a =  - \frac{d w}{dx^a} \frac{dx^a}{d\tau}= -\frac{d w}{d\tau}.
    \lb{ku}
\ee
Considering equation (\ref{w2}) we then conclude that,
\be { \left[ \frac{d w}{d\tau} \right] }_C=1, \lb{dwds} \ee
and
\be { \left[ k_a u^a \right] }_C = -1. \lb{ku2} \ee
This result allows us to write equation (\ref{dn2}) as below,
\be
  dN= ({d_{\scriptscriptstyle A}})^2 (1+z) d \Omega_0 \; n(y) \; dy.
  \lb{dn3}
\ee
This equation means that the local displacement at the rest-frame of
the galaxy can be rewritten in terms of the redshift, as follows,
\be
   dl= { \left[- u^a k_a \right] }_P \; dy=(1+z)dy.
   \lb{dl}
\ee
We can actually write equation (\ref{dn3}) completely in terms of $z$, with
$n$ and ${d_{\scriptscriptstyle A}}$ as functions of $z$, since
$z=z(y,\theta,\varphi)$. This is an important point to bear in mind, since
in the next pages we will often refer to the quantity $dN/dz$, but rarely
to $dN/dy$.

\subsubsection{Average Galactic Mass}\label{AGM}

Although the number density of sources is very important for testing 
cosmological models, so is the mass-energy density of the universe. Thus,
for completing our comparison of theory with observational data, we must at
some stage relate mass-energy density to the number density of sources. To
accomplish this we need to determine the average galaxy rest mass. At
present this is not possible to determine from observations. So, let us
indicate how it can be determined in principle, as well as discussing
briefly the rough estimate we use here, for the sake of illustrating our
general procedure. 

Let us define ${\cal M}_g$ to be the {\it average galaxy rest mass}.
Since galaxies of different morphological types may have masses varying
from $10^9$ to $10^{12}$ solar masses,\footnote{ \ These values are just
a rough estimate, based on Elmegreen (1998) and Sparke and Gallagher's
(2000) data, as dwarf galaxies may have smaller masses.} we may think
of ${\cal M}_g$ as the
simple average ${\cal M}_g= (1/V) \sum_{v=1}^{V} {\cal M}_v$, where $V$
is the number of galaxy morphological types being considered (ellipticals,
spirals, irregulars, etc) and ${\cal M}_v$ is the typical rest-mass
value for galaxies of a certain morphological class $v$. By ``typical''
we mean a simple average mass value. Note that this definition just
takes the masses measured in nearby galaxies and extrapolates them to
higher $z$.

Nonetheless, galactic class population abundances may vary with the
redshift due to galaxy evolution, that is, galaxies may change from
one morphological type to another in different epochs.
In addition, galaxy rest masses are
measured in the source's rest frame. Furthermore, we may consider that some
morphological types will be present only in certain redshift ranges,
and be absent in others. Therefore, in general ${\cal M}_v$ will be a function
of $z$, since the typical mass of a galaxy of a given type will vary
with redshift. ${\cal M}_v(z)$ will still be an average of sorts -- the average
mass of, for instance, spiral galaxies in a redshift bin $\Delta z$. What
${\cal M}_v(z)$ is would have to be determined observationally and/or
astrophysically, that is, with a combination of both observation and
theory.

Considering this, it is therefore more appropriate to define ${\cal M}_g$
by means of the following weighted average,
\be
   {\cal M}_g(z) = \sum_{v=1}^{V} P_v(z) {\cal M}_v(z),
   \lb{mg}
\ee
where $P_v$ is the {\it galactic morphology population fraction},
giving the abundance of each galactic type relative to the total number
of counted galaxies in each redshift range. This quantity is normalized by
\be \sum_{v=1}^{V} P_v(z_f)=1, \lb{p} \ee
where $z_f$ stands for a fixed redshift value, or a certain redshift
interval or bin $\Delta z$, to be determined observationally. 

Equation (\ref{mg}) as defined above does not account for or describe
galaxy mergers, as the basic information provided by it are the
proportions of each galaxy population in the sample's redshift range.
However, it does allow for mergers when ${\cal M}_g(z)$ is considered together
with the luminosity function (see \S \ref{lumi} below). In the simplest
possible case, we may assume just one morphological class ($V=1$), that,
of course, will not change with the redshift, and its mass will be equal
to the simple average of all galactic rest masses available. In this way
equation (\ref{mg}) reduces to
\be {\cal M}_g \approx 10^{11} {\cal M}_\odot. \lb{mg2} \ee
Note again that this factor is just a rough estimate based on data presented
by Sparke and Gallagher (2000, p.\ 204, 264; see their Table 5.1).

As stated above, we have written ${\cal M}_v(z)$ as a function of the
redshift $z$, since galaxy interactions, including mergers, will affect
the average galaxy masses of a given morphological type as the universe
evolves. Similarly, $P_v(z)$ also depends on $z$ -- as galaxies may
change from morphological type to another as we go from earlier epochs
to later ones. Typically, we find more spiral galaxies at higher
redshifts, and more ellipticals at lower redshifts, for instance. 
But how are we to determine ${\cal M}_v(z)$? In principle, this can
only be consistently done by using the galaxy luminosity functions for
each type of galaxy as a function of redshift (which are defined and
discussed below in \S \ref{lumi}), along some reliable measure of the
masses of the galaxies in the sample -- for instance the mass-to-luminosity
ratio for them (see \S \ref{mlr} below). Then, one could integrate the
luminosity multiplied by the mass-to-luminosity ratio and the galaxy
luminosity function over the luminosity, which is just the luminosity
density convolved with the mass-to-luminosity ratio (see eq.\ \ref{l}
below), for each morphological type in a given redshift interval, to
obtain the mass density contributed by that sample of galaxies.
This can then be divided by the integral over the galaxy
luminosity function itself for that particular
type in the same redshift interval, giving the number density of galaxies
of that type, to obtain the average galaxy mass for that sample. Perhaps,
we shall eventually be able to do this. However, at present there are
several problems. First of all, the luminosity functions diverge a the
faint end, and it is not clear how to choose reliably a faint-end cutoff.
There are a large number of dwarf galaxies, and we do not know where they
cease in the luminosity function. Furthermore, we cannot reliably determine
mass-to-luminosity ratios which will apply to the full range of galaxies
in the samples. 

Thus, converting galaxy number densities to mass-energy densities presents
us with a serious problem. We plan to investigate how this might be done
with some confidence in later work. For this paper, in order to illustrate
our general methods, we simply set the values of the ${\cal M}_v(z)$ on the
basis of the rough estimates for typical galaxies of given type $v$ as
provided by Sparke and Gallagher (2000) and Elmegreen (1998). These values
are given in \S \ref{testing}, where we apply our two consistency tests.
This is by no means adequate, but our focus in this paper is to illustrate
how our general framework can be used, when we have enough reliable
observational data. 

\subsubsection{Relativistic Density per Source}

The number density of {\it all} cosmological sources in their own rest
frames can be written as,
\be n(y)=\frac{\rho}{{\cal M}_g}, \lb{n} \ee
where $\rho$ is the local density as given by the right hand side of
Einstein's field equations. This presumes that the cosmological constant is
zero, and that there is no dark matter not associated with the sources we
are considering. If there is, then we can simply add those components to the
matter density of the sources to obtain $\rho$. Now, following Stoeger
\etal (1992, \S 3) we shall define the {\it relativistic density per source}
${\mu}(z)$ in terms of the {\it differential number counts} $dN/dz$ as
\be {\mu}(z) \equiv \frac{{\cal M}_g}{{{(d_{\scriptscriptstyle A})}^2}
d \Omega_0}
    \frac{dN}{dz}.
    \lb{mz2}
\ee
Thus, this quantity can, in principle, be determined observationally (Loh
and Spillar 1986; Newman and Davis 2000). Remembering equation (\ref{dn3}),
we can also write $\mu(z)$ in terms of the cosmological density $\rho$,
\be
{\mu}(z) = (1+z) \; \rho \; \frac{dy}{dz}. \lb{mz}
\ee
Interestingly, ${\mu}(z)$, or $dN/dz$, can be precisely determined
theoretically for any cosmological model ({\it e.g.,} for FLRW) and then
compared with what is measured observationally (see below). In addition,
one can clearly see that it is easier to find ${\mu}(z)$ from
observations by means of equation (\ref{mz2}), while it is easier to
find ${\mu}(z)$ from theory by means of equation (\ref{mz}). It goes
without saying, from what we have indicated above, that both ${\mu}(z)$
and $dN/dz$ depend on both cosmological evolution and astrophysical
number evolution.

The relationship between the observed differential counts and the total
count may be written as
\be { \left( \frac{dN}{dz} \right)}_0=J \left( \frac{dN}{dz} \right),
    \lb{j}
\ee
where $J$ is the {\it completeness parameter}, giving the percentage
of galaxies effectively observed in a galaxy redshift survey relative 
to the estimated total content of the sample. Thus,
\be 0 \le J \le 1, \lb{j2} \ee
and equation (\ref{mz2}) becomes,
\be {\mu}(z) = \frac{{\cal M}_g}{J{{(d_{\scriptscriptstyle A})}^2} d \Omega_0}
    { \left( \frac{dN}{dz} \right)}_0.
    \lb{mz3}
\ee
Notice that this equation is bolometric, since the observed differential
count on the right side is not restricted to any pass-band wavelength
observations. We can similarly write the observed relativistic density
of the universe ${\mu}_0(z)$ in terms of the observed differential
count as follows,
\be {\mu}_0(z) = \frac{{\cal M}_g}{{{(d_{\scriptscriptstyle A})}^2} d \Omega_0}
             { \left( \frac{dN}{dz} \right)}_0,
    \lb{mz0}
\ee
inasmuch as,
\be {\mu}_0(z) = J {\mu}(z).
    \lb{mzmz0}
\ee 

A final remark: although the equations above were written in terms of
the area distance, expressions for the other observational distance
definitions can be easily obtained by means of Etherington's (1933)
{\it reciprocity theorem}, valid for {\it any} cosmological metric. It
may be written as follows (see also Ellis 1971, pp.\ 153-156;
Schneider{ \it et al.\ }1992, pp.\ 110-114),
\be d_{\scriptscriptstyle L}=d_{\scriptscriptstyle A} { \left( 1+z \right)
    }^2=d_{\scriptscriptstyle G} \left( 1+z \right),
    \lb{recip}
\ee
where $d_{\scriptscriptstyle L}$ is the usual {\it luminosity distance}, and
$d_{\scriptscriptstyle G}$ is the {\it galaxy area distance}.\footnote{ \ In
the literature $d_{\scriptscriptstyle G}$ is also known as {\it angular size
distance}, {\it effective distance}, {\it transverse comoving distance}, and
{\it proper motion distance} (Scott \etal 2000, p.\ 644; Ribeiro 2001,
footnote 9, p.\ 1711).} The area distance is, in principle, an observationally
determinable quantity, as its definition is given by (Ellis 1971, p.\ 153),
\be { \left( d_{\scriptscriptstyle A} \right) }^2=\frac{dS_0}{d \Omega_0},
    \lb{da2}
\ee
where $dS_0$ is the cross-sectional area at point $P$ of a bundle of null
geodesics converging towards the observer (see figure \ref{fig1}). So,
if, by means of some astrophysical model, we are able to infer the intrinsic
cross-sectional area $dS_0$ of some object, and measure its subtended
solid angle $d \Omega_0$, then $d_{\scriptscriptstyle A}$ can be found.
Notice, however, that the cross-sectional area $dS_0$ is defined in the
rest-frame of the source.

\subsection{Galaxy Luminosity Function}\lb{lumi}

The galaxy {\it luminosity function} $\phi$ is defined to be the {\it
local} number density of {\it all} galaxies with absolute luminosity
between $L$ and $L+dL$ at redshift $z$ (Weinberg 1972, p.\ 452; Peebles
1993, p.\ 119; Peacock 1999, p.\ 399),
\be \frac{dN(z, L)}{dV} = \phi ( \ell ) d \ell,
    \lb{lm}
\ee
where
\be \ell \equiv  L/L_\ast, \lb{ell} \ee
$L_\ast$ is the luminosity scale parameter, and $dV$ is the {\it proper
volume element} along the null cone, at some redshift, where the
luminosity function is determined.\footnote{ \ In the past there was
sometimes confusion or unclarity about which volume definition was
being used in a given survey when one studies the literature on
galaxy luminosity function determination. The proper volume was
initially the one mostly used, but when the surveys started to probe
at redshifts deeper than $0.1$ or $0.2$, most observers switched to
{\it comoving volume} usage instead, which is now the prevalent
volume definition among observers (Chris Impey 2002, 2003,
private communications). Theoreticians, on the other hand, often adopt the
proper volume definition. Therefore, in order to have a clear link
between theory and observation, in this paper we shall adopt
the proper volume as our volume definition for the luminosity function
determination, unless stated otherwise. Wherever another volume definition
becomes necessary in order to compare our theory with observational data,
we will employ an appropriate conversion factor (see \S \ref{vols} below).}
Integration of $\phi$ above a certain luminosity threshold $L$ defines
the {\it selection function} $\psi$ (Peebles 1993, p.\ 214),
\be
   \psi \left[ \ell(z)\right] =\int_{\ell(z)}^{\infty} \phi ( \ell ) d \ell,
   \lb{psi}
\ee
while summing $\phi$ over all luminosities yields, 
\be
    \Psi(z) =\int_0^{\infty} \phi ( \ell ) d \ell.
    \lb{Psi}
\ee
This simplified formulation is an idealization. There is no universal galaxy
luminosity function. In actuality, there are, instead, different 
luminosity functions for different types of galaxies and for different
observational band-widths, and these evolve with redshift. Here we shall
continue to let the unindexed luminosity function represent the distribution
of all the galaxies, and designate the luminosity functions of certain
classes of galaxies in certain observational band-width with subscript and
superscript indices, respectively.

The introduction of the luminosity and selection functions means that
in order to be detected and counted a galaxy must have a minimum flux,
or apparent luminosity $F$, which is related to its intrinsic luminosity
$L$, at some redshift $z$, by the well-known expression
\be F=\frac{L(z)}{4\pi {\left( d_{\scriptscriptstyle A} \right) }^2
    { \left( 1+z \right) }^4}, 
    \lb{f}
\ee
for the case of bolometric flux. For a specific bandwidth range $W$,
the apparent luminosity is further restricted to (Ellis 1971, p.\ 161;
Ribeiro 2002),
\be F_{\scriptscriptstyle W}= \frac{L(z)}{4\pi {\left(
       d_{\scriptscriptstyle A} \right) }^2
    { \left( 1+z \right) }^3} \int_0^\infty W(\nu) {\cal J} \left[ \nu (1+z)
    \right] d\nu,
    \lb{fw}
\ee
where $W(\nu)$ defines the spectral interval of the observed flux (for
instance, the standard UBV system) and ${\cal J}$ is the {\it source
spectrum function}, that gives the proportion of radiation emitted by
the source at some rest-frame frequency, but redshifted and detected by the
observer as $\left[ \nu (1+z) \right]$ (Ellis 1971, p.\ 161; Ribeiro
2002).

These two expressions for $F$ take into account relativistic effects. If
those effects were to be left out, we would have to use their Euclidean
counterparts, suitable for a non-expanding universe approximation. In this
case the bolometric flux is given by,
\be
    F = \frac{L}{4 \pi { \left( d_{\sssty L} \right) }^2},
    \lb{feuc}
\ee
which could also be found by using the reciprocity theorem (\ref{recip})
in equation (\ref{f}). Similarly, the Euclidean equation for
$F_{\scriptscriptstyle W}$ can also be found. Using the equation above
for determining the luminosity function will strongly affect the galaxy
counts, as the lower limit of the integral in equation (\ref{psi}) will
be changed (Peebles 1993, p.\ 214).

The incompleteness of the sample is indicated by the lower limit of the
integral in equation (\ref{psi}), since faint galaxies with intrinsic
luminosity below $L(z)$ will not be counted. If we now suppose that
the observationally determined luminosity function for brighter
galaxies may also be used to account for those faint galaxies, we
may then write the completeness parameter as,
\be J(z) = 1 - \frac{1}{\Psi(z)} \int_0^{\ell(z)} \phi ( \ell )
    d \ell,
    \lb{j3}
\ee
which agrees with condition (\ref{j2}), as it should. In addition,
the selection function may be written as,
\be 
  \psi \left[ \ell(z)\right] = J(z) \Psi(z).
 \lb{pP}
\ee

In order to provide a relativistic version for equation (\ref{lm}), we
must link the definitions above to equation (\ref{dn3}). However, it is
clear from their definitions that the only difference between $n$, as
in equations (\ref{dn}), (\ref{dn2}) and (\ref{dn3}), and $\phi$ is
that the latter restricts the rest-frame volume count to a luminosity
range. Considering this, we can only conclude that,
\be
  n(y) = \Psi \left[ z(y) \right].
  \lb{Psi3}
\ee

Notice that this equation assumes that {\it both} quantities above
are measured relative to proper volume elements. If, however, $\Psi$
is being observationally determined by means of another volume definition,
like the comoving volume, an appropriate conversion factor must be
introduced in order for equation (\ref{Psi3}) be correct (see \S
\ref{vols} below). 

In addition, the functional relationship between the redshift $z$ and
the affine parameter $y$ remains unknown in the general case, and must
be determined by solving the field equations (in particular, the null
Raychaudhuri equation). Thus, once a cosmological metric is assumed or
specified, it can be determined. In other words, to fix the function
$y=y(z)$ we require a cosmological solution of Einstein's field
equations.\footnote{ \ In general this function will also depend on
$\theta$ and $\varphi$.} Of course, in practice we may be using some null radial
coordinate other than the affine parameter, say $r$. Then we would need
$r = r(z)$ or its inverse relationship. Below, in dealing with the
Einstein-de Sitter cosmological model, we shall use an $r$ which is
closely related to the radial coordinate which is often used in
describing FLRW models.

We can write the relativistic number counts in terms of the luminosity
function by substituting the result above into equation (\ref{dn3}). The
result yields,
\be
   dN={\left( d_{\scriptscriptstyle A} \right) }^2 (1+z)d \Omega_0
      \Psi(y) dy.
   \lb{dn5}
\ee
Hence, the observed differential number counts for sources whose intrinsic
luminosity are above the threshold $L(z)$ becomes,
\be { \left[ \frac{dN (z,>L)}{dz} \right] }_0 ={ \left(
     d_{\scriptscriptstyle A} \right) }^2 (1+z) d
    \Omega_0 \; \psi \left[ L(z)\right] \frac{dy}{dz},
    \lb{lm3}
\ee
while the total count yields,
\be \frac{dN(z)}{dz} = { \left( d_{\scriptscriptstyle A} \right) }^2
    d \Omega_0 (1+z) \Psi \frac{dy}{dz}.
    \lb{lm2}
\ee
The observed relativistic density per source in the universe will then
be given by,
\be
   {\mu}_0 \left[ z(y) \right] = {\cal M}_g (1+z) \psi \frac{dy}{dz},
   \lb{mz04}
\ee
which implies that,
\be
    \rho = {\cal M}_g \Psi,
    \lb{rho2}
\ee
if we assume the cosmological constant is zero and that all dark matter is
associated with the sources. Otherwise, we would have to write 
$\rho_m = {\cal M}_g\Psi,$ where $\rho_m$ represents only the matter associated
with the sources themselves, not vacuum energy nor, for instance, matter 
distributed differently from the observed sources. 
These equations can be completely determined in terms of the redshift
once the function $y=y(z)$ is found for a specific cosmological
spacetime, allowing us to eliminate the affine parameter.

For bolometric counts the lower limit $L(z)$ is given by equation
(\ref{f}), while for counts in the bandwidth $W$ we must use equation
(\ref{fw}). If observations are in a limited bandwidth $W$, then the
observed relativistic density per source of the universe must be written as,
\be {\mu}_0^{\scriptscriptstyle W}(z)= \sum_{v=1}^V P^{\sssty W}_v(z)
    {\cal M}_v(z) J^{\scriptscriptstyle W}_v(z) (1+z) \Psi^{\sssty W}_v (z)
    \frac{dy}{dz},
    \lb{mz5}
\ee
where
\be \Psi^{\sssty W}_v (z)=\int_0^\infty \phi_v^{\scriptscriptstyle W}
    ( \ell ) d \ell, 
    \lb{Psi2}
\ee
and $J^{\scriptscriptstyle W}_v(z)$ is the completeness parameter
determined by means of the bandwidth luminosity as given by equation
(\ref{fw}), but with a luminosity function for each galaxy morphological
type $v$ and at different wavebands $W$. Notice that since the luminosity
function conveys observational information about the number of sources
at each redshift bin, it will effectively take into account the possible
{\it galaxy mergers} taking place in those ranges, with the morphological
population fraction $P^{\sssty W}_v(z)$ conveying the observational
relative abundance of each morphological type in the same redshift
ranges, and for each observing bandwidth $W$. In other words, number
evolution and cosmological evolution are contained in the luminosity
function, but number evolution will also be reflected in ${\cal M}_g(z)$.

Remembering equation (\ref{mzmz0}), the observed relativistic density
per source
can, for the best possible observations, be at most equal to its
theoretical counterpart, ${\mu}_0(z) \le {\mu}(z)$. However, if we
now suppose the existence of some form of dark matter, we should rewrite
equation (\ref{mzmz0}), since the total theoretical value of ${\mu}$
should be given by the following expression,
\be
   { \mu}(z)= \frac{ {\mu}_0(z)}{J} + {\mu}_d,
   \lb{mzm0}
\ee
where ${\mu}_d$ accounts for the possible {\it dark matter
component}. Therefore, from now on all theoretical quantities will be
decomposed into an observational part, and a possibly unseen, differently
distributed, component.

According to this definition, ${\mu}_d$ does not include dark matter
galactic halos, as this is already included in the rest mass
determination of each galaxy. So, ${\mu}_d$ may be baryonic intracluster
dark matter, or non-baryonic dark matter, or vacuum energy, or still
possibly, quintessence.

In writing ${\mu}_0(z)$ we have also to consider that it is composed
by summing up the individual contributions of the different bandwidths $W$,
$W'$, etc, where the observations are actually made. Therefore, we have
that,
\be {\mu}_0(z) = {\mu}_0^W+{\mu}_0^{W'}+{\mu}_0^{W''}+
    \ldots 
    \lb{mz6}
\ee
This equation is valid provided that the different bandwidths $W$,
$W'$, etc, do not overlap, otherwise there will be double counting.
However, in general we may expect significant overlap of galaxy counts
over different bandwidths, meaning that equation (\ref{mz6}) will convey
incorrect results due to overcounting of the galaxies we observe. In order
to avoid this situation, let us rewrite equation (\ref{mz6}) as follows,
\be
   {\mu}_0(z) = \sum_{{\scriptscriptstyle W}=1}^{\scriptscriptstyle
   W_f} a_{\sssty W}(z) {\mu}_0^{\sssty W}(z),
   \lb{mz7}
\ee
where
\be a_{\sssty W}(z) =1, \; \; \; \; \mbox{for} \; \; \; \;  W=1,
  \lb{aw}
\ee
and
\be a_{\sssty W}(z)=b_{\sssty W}(z) < 1, \; \; \; \; \mbox{for} \; \; \;
    \; W>1.
  \lb{bw}
\ee
Here $b_{\sssty W}(z)$ is the fraction of galaxies in waveband $W>1$
which are not counted in wavebands $1,2,\ldots,(W-1)$.

For later usage it is convenient to define the {\it luminosity density}
as being given by (Peebles 1993, p.\ 120),
\be j = \int_0^\infty L \; \phi ( \ell ) d \ell.
    \lb{l}
\ee
This is the average luminosity of the sources per unit of proper or
comoving volume.

\subsection{Observed Average Densities}

The {\it observed average density of the universe} is defined as follows,
\be \langle \rho \rangle \equiv \frac{{\cal M}_g N}{V_0}, \lb{ad} \ee where
$V_0$ is the {\it total observed volume}. As discussed in Ribeiro (2001,
and references therein), equation (\ref{ad}) does not define an unique
quantity, but at least three, as the observational volume can be
constructed with any of the three distance definitions given by the
reciprocity theorem (\ref{recip}). Consequently, we have that
\be V_0 \equiv \frac{4}{3} \pi {d_0}^3,
    \lb{v0}
\ee
where $d_0$ may be any of the three distances, $d_{\scriptscriptstyle L}$,
$d_{\scriptscriptstyle A}$, or $d_{\scriptscriptstyle G}$.\footnote{ \
By means of the observed proportionality between redshift and distance,
{\it i.e.}, the Hubble law, in FLRW cosmologies one can also define the
{\it redshift distance}, given by $d_z \equiv cz/H_0$, where $c$ is the
light speed and $H_0$ is the Hubble constant. Therefore, a fourth
distance definition is also available in the standard model, as well as
another total observed volume. For non-FLRW metrics, one can only have a
similarly defined $d_z$ if an equivalent to the Hubble parameter is also
defined in the chosen geometry ({\it e.g.}, Ribeiro 1994; Pompilio and
Montuori 2002).}\lb{note5} As at the observer's position $C(q)$ the
redshift is zero, these definitions must imply that 
\be \lim_{z \rightarrow 0} \langle \rho \rangle = \rho_0 .
    \lb{avr}
\ee
Notice that the observational quantities so far used, namely % ${\mu}(z)$,
$n$, and $\phi$, are {\it not} average densities, but local densities,
defined at the source's rest frame. 

To determine $\langle \rho \rangle$ the key quantity to be found is $N$.
Bearing in mind the results previously derived, the number of observed
sources above a certain luminosity $L(z)$ is given by,
\be
  N_0 (z,> L)= \sum_{{\sssty W}=1}^{\sssty W_f} a_{\sssty W}(z)
       N_0^{\sssty W} (z) = \sum_{{\sssty W}=1}^{\sssty W_f}
       a_{\sssty W} \int_0^y { \left( d_{\sssty A} \right) }^2
       (1+z) \; d \Omega_0 \; {J^{\sssty W}} \; \Psi^{\sssty W} \; dy,
  \lb{N0}
\ee
while the overall number of sources of any luminosity is,
\be N(z) = \sum_{{\sssty W}=1}^{\sssty W_f} a_{\sssty W}(z)
           N^{\sssty W} (z) = \sum_{{\sssty W}=1}^{\sssty W_f}
           a_{\sssty W} \int_0^y { \left( d_{\sssty A} \right) }^2
          (1+z) \; d \Omega_0 \; \Psi^{\sssty W} \; dy.
  \lb{N}
\ee

We can now write the observed and total average densities of the
universe in terms of the previously defined quantities. The expressions
yield,
\be
  { \langle \rho \rangle }_0 = \frac{3}{4 \pi
  { \left( d_0 \right) }^3 } \sum_{{\sssty W}=1}^{\sssty W_f} a_{\sssty W}
  (z) \sum_{v=1}^V {P^{\sssty W}_v(z) {\cal M}_v(z)} \int_0^y { \left( d_{\sssty A}
  \right) }^2 (1+z) \; d \Omega_0 \; \psi_v^{\sssty W} \; dy,
  \lb{avg}
\ee
and
\be
  \langle \rho \rangle  = { \langle \rho \rangle }_d + \frac{3}{4 \pi
  { \left( d_0 \right) }^3 } \sum_{{\sssty W}=1}^{\sssty W_f} a_{\sssty W}
  (z) \sum_{v=1}^V {P^{\sssty W}_v(z) {\cal M}_v(z)} \int_0^y { \left( d_{\sssty A}
  \right) }^2 (1+z) \; d \Omega_0 \; \Psi_v^{\sssty W} \; dy,
  \lb{avg2}
\ee
where ${ \langle \rho \rangle }_d$ is the possible dark matter component.
In redshift bins where $\phi$ varies little with $z$, we have the 
following result,
\be
   { \langle \rho \rangle }_0= J { \langle \rho \rangle } - { \langle
   \rho \rangle }_d.
 \lb{54}
\ee

It is clear from equations (\ref{avg}) and (\ref{avg2}) that building an
average density from observations is a very complex task which depends on many
choices, like the appropriate distance definitions to be used, and a still
to be determined function $y=y(z)$ (if we do not assume a cosmology to begin
with), which is certain to be non-linear.
In addition, the complexity and high non-linearity of $\langle \rho
\rangle$ shows the inherent difficulties of the task of observationally
ascertaining whether or not the universe is smooth when one follows a
fully relativistic approach of this problem (Ribeiro 2001). 

\subsection{Mass-to-Luminosity Ratio}\label{mlr}

The {\it mass-to-luminosity ratio} at some redshift depth is defined as,
\be {\cal M} / L  \equiv \frac{\rho}{j}={\cal M}_g \frac{\Psi}{j}, 
    \lb{ml}
\ee
while its observable counterpart is given by,
\be
   { \left[ {\cal M} / L \right] }_0 = {\cal M}_g \frac{\psi}{j}.
   \lb{ml0}
\ee
If the {\it average luminosity of the universe} is similarly defined as
the average density $\langle \rho \rangle$, we have that,
\be
   \langle j \rangle = \frac{L_g N}{V_0},
   \lb{lg}
\ee
where $L_g$ is the {\it average galaxy luminosity}, given by
\be 
   L_g(z) = \sum_{v=1}^V P_v(z) L_v(z).
   \lb{lg2}
\ee
Here $L_v$ is the typical luminosity value for galaxies of a certain
morphological type $v$, and $P_v$ is the galactic fractional
population class (see eqs.\ \ref{mg} and \ref{p}).

{From} the equations above, the {\it average mass-to-luminosity ratio}
at some redshift depth is given by,
\be \langle {\cal M}/L \rangle = \langle \rho / j \rangle =
    \frac{ \langle \rho \rangle }{ \langle j \rangle}=
    \frac{ { \langle \rho \rangle}_0 }{ { \langle j \rangle }_0 }=
    \frac{ \sum_{v=1}^V P_v (z) {\cal M}_v (z)}{ \sum_{v=1}^V P_v(z) L_v (z)}.
    \lb{avml}
\ee

\section{Luminosity Function Data}\lb{luminosity}

The discussion presented so far has focused on galaxy number counts,
which, aside from the redshift, is one of the two observational quantities
(observer area distance, or luminosity distance is the other) which
determine the geometry and mass-energy distribution of the Universe
(assuming spherical symmetry - if this is not assumed, then we need two
other quantities, cosmological proper motions and null shear parameters -
see Ellis \etal 1985). Thus, the ideal way to determine the mass-energy
distribution and the geometry would be by solving the field equations
directly with redshift, number count and observer area distance data.
However, as already mentioned, our ignorance of luminosity and number
evolution, and adequate model of which would have to be assumed, is an 
obstacle to implementing this in a compelling way. Because of this, and
since the luminosity functions available in the literature presuppose a
definite FLRW cosmology, the first and
easiest sensible application of the general theory we have summarized
above is to test luminosity functions for consistency with observational
relationships determined by the cosmologies they have assumed. In other
words, in any specific cosmological metric we
need to make both sides of the equation \begin{center} [geometric
cosmological model] $\leftrightarrow$ [parameterized data] \end{center}
consistent, as we will be dealing either with quantities which are
determined only observationally ({\it e.g.}, $\phi$), or with 
theoretical quantities derived from the chosen cosmological model
({\it e.g.}, ${\mu}$, $N$), or both ({\it e.g.}, $dN/dz$, ${\cal M}/L$)
(Stoeger 1987, p.\ 299).

The luminosity function provides an excellent observational tool for
testing cosmological models, as it conveys information on how the local
number of galaxies $n$ vary for different types of galaxies, its
environmental dependence, how $n$ changes with the cosmic epoch (see
eq.\ \ref{Psi3}), and also a possible indirect estimate of the amount of dark
energy present. The luminosity functions also provide a statistical
description of the galaxy population. The functional form of $\phi$ and
its parameter values are extracted from magnitude-limited redshift
surveys of galaxies by means of various statistical techniques (Peebles
1993, p.\ 119; Binney and Merrifield 1998, p.\ 162; Peacock 1999,
\S 13.3). Here, however, we shall only be interested in collecting
from the literature the shape, normalization, and the luminosity scale
of $\phi$ for the purpose of using it in our equations for the mass-energy
density and the relativistic mass-energy density per source, and 
checking that the results are consistent with the cosmology assumed in
calculating the luminosity functions. Some of the information contained
in the luminosity functions is tied to the galaxy morphological or
spectral class (see below), which means that at least a simplified
morphological division must be assumed.

As discussed by Chen {\it et al.\ }(2001), various deep redshift
surveys have yielded consistent measurements of the luminosity function
for galaxies at $z \apl 0.75$ (Marzke {\it et al.\ }1998; Lin {\it et al.\
}1999; Cole \etal 2001), and this is precisely the redshift depth we are
interested in here (see \S \ref{intro}). Impey and Bothun (1997) have also
compiled valuable observational data on $\phi$. Since these studies yield
recent data on $\phi$, they seem reliable for giving the luminosity
functions for this redshift range.

\subsection{Morphological Types}\lb{morphology}

As it is well known, galaxies have different morphologies, which vary 
from ellipticals and spheroidals to spirals and irregularly
shaped galaxies (Binney and Merrifield 1998, ch.\ 4; Elmegreen 1998,
ch.\ 2; Sparke and Gallagher 2000, \S 1.3). Redshift surveys dealing
with thousands of galaxies tend, however, to automatically classify
galaxies according to their spectral energy distributions (Lin \etal
1999, p.\ 539; Cole \etal 2001, \S 5.3), or by means of a simple
three-type scheme, as E/S0, spiral, and irregular/peculiar galaxies
(Marzke \etal 1998). It is important to mention that spectral type
classifications are not necessarily straightforwardly related to the
galaxy types of the Hubble diagram, and its extensions, and, since
many surveys do adopt such a scheme, for the purposes of this paper it
seems enough to follow Lin's \etal (1999) spectral morphological types.
So, from now on we will assume $V=3$ (see eq.\ \ref{mg}), for three
loose classification types: $v=1$ stands for {\it early type galaxies},
roughly E/S0; $v=2$ accounts for {\it intermediate type galaxies},
something like Sa-Sb; and $v=3$ represents {\it late type galaxies},
which would be Sc-I types. Notice, again, that this equivalence between
the galactic spectral energy distributions and morphological types is
not strict, and, therefore, should not be taken as giving a precise
classification.

The classification scheme above is not unique. For instance, with the
data presented by Impey and Bothun (1997) one could think of only two
galactic types: $v=1$ for galaxies in clusters, which tend to be dominated
by spirals, and $v=2$ for field galaxies, which are usually ellipticals.
One could even think of other classification schemes, like luminous galaxies
($v=1$), low surface brightness galaxies ($v=2$), and luminous giants
galaxies ($v=3$), or still $v=1$ for red galaxies, and blue galaxies as
$v=2$. Perhaps, combinations of all these schemes are possible, although
in such a case the problem of overcounting could become more acute.

Nevertheless, whatever classification scheme is chosen when applying the
theory developed above, the bottom line is that the adopted morphological
types will be dependent basically on the sample, or on the sampling
method, used in the chosen redshift survey. In most cases it will probably
not involve strict identification with the galaxy morphologies of the Hubble
diagram. So, from now on the term ``morphological type'' will have a
broad meaning, referring to some sort of classification obtained by 
differentiating the galaxies contained within a certain sample. Such a
classification can be based either on spectra, or on the galactic
shapes as given by the Hubble diagram, or on any other galactic
observational feature considered important by some sampling method.

\subsection{Shape}

In finding the shape of the luminosity function, most, if not all,
recent redshift surveys fit their data by using Schechter's (1976)
elegant form, which summarized earlier findings by Zwicky and Abell that
the luminosity function is a broken power law, $\phi \sim L^{-1.5}$ at
$L < L_\ast$, bending to $\phi \sim L^{-3}$ at the bright end (Peebles
1993, p.\ 119). {\it Schechter's luminosity function} is given by,
\be
     \phi(\ell)= \phi_\ast \; \ell^\alpha e^{- \ell},
     \lb{sch}
\ee     
where $\phi_\ast$ characterizes the space density of galaxies, and $\alpha$
is the asymptotic slope of the faint end of the luminosity function. The
luminosity scale $L_\ast$ is given in $\ell$ (see eq.\ \ref{ell}), and
tells us the luminosity above which galaxies are rare. These three
constants are all observationally determined parameters.

The power law index $\alpha$ for the faint end slope is part of the
shape determination of the luminosity function, and, according to the
latest observations, it does change with the morphological class of the
surveyed galaxies, whether they are field galaxies, dwarfs, giants,
spiral-rich galaxy clusters, luminous galaxies or low surface brightness
galaxies (Impey and Bothun 1997). The analytical form of equation
(\ref{sch}), however, does not appear to vary with galactic properties.
Only the values of the parameters do.

In the early days of the work on the luminosity function,
all parameters were fitted to data only for local galaxies, where
evolution is not considered to be important. However, as the surveys went
deeper and deeper it became clear that possible luminosity evolution
had to be taken into account. This can be done by allowing some of the
fitted parameters to vary with the redshift (Lin \etal 1999), or by
comparing their values at different redshift bins (Chen \etal 2001). 
Obviously, a great deal of work is needed to model luminosity
and number evolution accurately, such as that pioneered by Totani,
Yoshii, and their collaborators (Totani and Yoshii 2000; Totani \etal
2001). We are not concerned in this paper with doing this. In order to
illustrate how our framework can be used, we provisionally accept the
rough evolutionary models assumed in the CNOC2 survey and test them
for cosmological consistency (see below).

\subsection{Normalization and Luminosity Scale}

Once the shape is determined, which includes determination of the
faint-end slope, that is, the $\alpha$ parameter, we are still left
with two unknowns, the normalization factor $\phi_\ast$ and the
luminosity scale $L_\ast$. Various determinations of $\phi_\ast$
show it not to be very sensitive to different FLRW models, even to
those with non-zero cosmological constant, meaning that up to
$z \approx 1$ these measurements appear to be robust. The same
seems also true for $\alpha$ and $L_\ast$. The major differences
in these quantities seem to come from measurements of different
morphological types and in different bandwidths (Binney and Merrifield
1998, p.\ 167; Lin \etal 1999; Cole \etal 2001). So, the present data
seems to indicate only a weak dependence of these parameters on the
cosmological models (Cross \etal 2001).

Redshift surveys usually present the fitted $L_\ast$ parameter in terms
of its {\it characteristic absolute magnitude} $M_\ast$. So, in order to
use these we have to rewrite equation (\ref{lm}) for the luminosity
function as the number density of all galaxies between absolute magnitudes
$M$ and $M+dM$. For Schechter's form (\ref{sch}) that becomes (Binney and
Merrifield 1998, p.\ 163),
\be
    \phi(\ell) \; d \ell=
    \phi_\ast \; \ell^\alpha e^{- \ell} d \ell =
    \left( 0.4 \ln 10 \right) \phi_\ast 10^{0.4 \left( 1+\alpha
    \right) \left( M_\ast - M \right) } \exp \left[ -10^{0.4 \left( M_\ast
    - M \right)} \right] dM 
    =\phi(M) \; dM,
    \lb{sch2}
\ee
since the definition of absolute magnitude implies that $\ell=10^{0.4(
M_\ast - M)}$.

\subsection{The CNOC2 Redshift Survey Data}\lb{cnoc2}

We have chosen to extract data to be used here from the {\it Canadian
Network for Observational Cosmology Field Galaxy Redshift Survey} (CNOC2),
presently the largest such a sample at intermediate redshifts. The CNOC2 sample
for luminosity function (LF) determination contains over 2000 galaxies in the
range $0.12 < z < 0.55$, and with apparent magnitude in the red band
within $17.0 < R_c < 21.5$. Their fitting parameters were confirmed at
$z \sim 0.75$ (Lin \etal 1999; see also Chen \etal 2001). The authors
also claim their fitting to be valid with little difference for $0 < z < 1$.
CNOC2 data were obtained and fitted in three bandwidths, $B_{\sssty AB}$, $R_c$,
and $U$, and are available for FLRW models with $q_0=0.5$ and $q_0=0.1$.
Therefore, by providing data in the redshift range we are interested in
for different bandwidths, and presupposing the Einstein-de Sitter
cosmological model, the CNOC2 survey is the most suitable to exemplify
how we can use this theoretical-observational framework for testing the
consistency of the LF's with the assumed cosmology, for determining how
much dark matter there is distributed differently than what is assumed
associated with galaxies, and for eventually determining the range of
validity of simplifying assumptions at low redshifts, among other
things.

The required data for the CNOC2 LF's assuming EdS universe are
summarized in the tables \ref{table1}, \ref{table2}, and \ref{table3},
where $v$ denotes the three morphological types in which the sample
population is divided: early, intermediate and late spectral types
(see \S \ref{morphology} above), and $P_v$ the proportion of each
galaxy population as compared to the whole sample. This is a fixed
value in all CNOC2 redshift range.  $M_\ast^\prime$ and $Q$ are two
constants of the linear equation for the characteristic luminosity
scale evolution (Lin \etal 1999, \S 3.2),
\be
   M_\ast (z)= M_\ast^\prime - Q (z-0.3),
   \lb{evol}
\ee
and $\phi_\ast$ is in units of h$^3$ Mpc$^{-3}$ mag$^{-1}$.

In computing these parameters of the LF from the CNOC2 survey, it is
important to note that Lin \etal (1999) also adopted a model for
galaxy number density evolution and assumed that the
parameter $\alpha$, which governed the shape of the LF at the faint
end, does not change with redshift for a given morphological type.
They were also careful to consider surface density selection effects,
and estimate the incompleteness of their sample, especially at the
higher end of their redshift range. Finally, they performed various
consistency checks to make sure that the LF parameters derived 
adequately represent the data for the redshifts indicated. It is
reassuring that Cohen (2002) finds that her results for the Caltech
Faint Galaxy Redshift Survey agree very well with the CNOC2 results,
including for the faint-end slope of the LF's. Similarly, the results
from the Deep Groth Strip Survey (Im \etal 2002) are in agreement with
the CNOC2 results, especially for the luminosity evolution parameter
$Q$. The one discrepancy that has emerged is that the significant
number evolution for late-type galaxies in the CNOC2 results is not
found by either Cohen (2002) or Im \etal (2002). Im \etal (2002) speculate
that this may be due to the fact that the CNOC2 samples were selected
purely by colors. This means at low redshifts ($z<0.4$), where the
difference in colors for various galaxy types is small, photometric
errors can easily bump blue galaxies into the red-galaxy samples. Other
than this, the CNOC2 LF results seem well supported for this redshift
range.

The determination of the LF's with CNOC2 data employed the comoving
volume instead of the proper volume (Huan Lin 2002, private communication).
Therefore, to be usable for us here we must apply an appropriate volume
conversion factor to the results listed in tables 1-3 (see
\S \ref{vols} below).

\section{CNOC2 and Einstein-de Sitter}\lb{flrw}

As already mentioned, we now want to use our observational-theoretical
framework first to test the consistency of the CNOC2 LF results with
the cosmological model it assumes, which is EdS. This has the added
advantage that EdS is mathematically the simplest of the FLRW models.
Furthermore, it is a frequently used case in the LF's available in
the literature. The extension of this type of analysis to other more
general FLRW cosmologies, including those with a non-zero cosmological
constant, are the subject of a forthcoming paper (Stoeger and Ribeiro
2003).

\subsection{The Metric}
 
Let us write the EdS metric as follows $(c=G=1)$,
\be
  ds^2=dt^2-a^2(t) \left[ dr^2 + r^2 \left( d \theta^2 + \sin^2 \theta
       d \varphi^2 \right) \right],
  \lb{eds}
\ee
where $a(t)$ is the scale factor, given by
\be 
  a(t)= { \left( t + \frac{2}{3H_0} \right) }^{2/3},
  \lb{factor}
\ee
$H_0$ is the Hubble constant, and the local density is
\be
   \rho= \frac{1}{6 \pi a^3(t)}.
   \lb{densi}
\ee
The solution of the past light cone equation, $dt/dy=-a(t)dr/dy$, may be
written as,
\be
  3 { \left[ t(y) + \frac{2}{3H_0} \right] }^{1/3}={ \left(
  \frac{18}{H_0} \right) }^{1/3}-r(y),
  \lb{plc}
\ee
which allows us to write equation (\ref{factor}) along the null cone
parameterized by $r$, as follows,
\be
   a[t(r)]= \frac{1}{9} { \left[ { \left( \frac{18}{H_0} \right) }^{1/3}
   - r \right] }^2.
   \lb{factor2}
\ee

Detailed calculations on this spacetime along the lines followed in this
article are widely available (see, for instance, Ribeiro 1995, 2001).
Therefore, here we will only summarize some important results required in
what follows.

\subsection{Observables}

Equations (\ref{factor}) and (\ref{factor2}) allow us to obtain the
function $z=z[r(y)]$, that is, an expression for the redshift where
the affine parameter becomes implicit. In EdS it may be written as,
\be
 1+z=\frac{a(t=0)}{a(t)} = { \left( \frac{18}{H_0} \right) }^{2/3}
    { \left[ { \left( \frac{18}{H_0} \right) }^{1/3}- r \right] }^{-2}.
    \lb{z2}
\ee
Applying equation (\ref{da2}) to the metric (\ref{eds}) gives us the
area distance in this cosmology, which, by means of equation (\ref{z2}),
yields,
\be 
   d_{\sssty A} = r a(t)= \frac{2}{H_0} \left[ \frac{ \sqrt{1+z} -1}
     { { \left( 1+z \right) }^{3/2}} \right],
   \lb{da3}
\ee
since along the past null cone the scale factor may be written as,
\be
   a= { \left( \frac{18}{H_0} \right) }^{2/3} \frac{1}{9 { \left( 1+z
   \right) }}.
   \lb{scalefactor}
\ee
In EdS geometry, for comoving sources, $u^b=\delta^b_0$, we have that,
\be
   -k^bu_b=-k^b g_{0b}=-k^0=-dt/dy=a(dr/dy),
   \lb{ku4}
\ee
which substituted into equation (\ref{dn}), and also remembering
equation (\ref{n}), gives us the total relativistic density per source
(\ref{mz2}) in the Einstein-de Sitter cosmology, as written below,
\be
  {\mu}(z)= \frac{{\cal M}_g}{ { \left( d_{\sssty A} \right) }^2 d\Omega_0 }
               \frac{dN}{dz} = \frac{3H_0}{8 \pi} \sqrt{1+z}.
  \lb{meds}
\ee
{From} these expressions, the results below readily follow.
\be
  \frac{dN}{dz}= \frac{3 d \Omega_0}{2 \pi H_0 {\cal M}_g}
                 \frac{ { \left( \sqrt{1+z} - 1 \right) }^2}{ { \left(
		 1+z \right) }^{5/2}},
  \lb{dndz}
\ee
\be
   N(z)= \frac{d \Omega_0}{\pi H_0 {\cal M}_g} { \left( 1 -
         \frac{1}{\sqrt{1+z}} \right) }^3,
   \lb{nz}
\ee
\be
   \rho(z)= \frac{3 {H_0}^2}{8 \pi} { \left( 1+z \right) }^3.
   \lb{rhoz}
\ee 
Finally, with either equations (\ref{mz}) and (\ref{meds}), or
(\ref{dn3}) and (\ref{dndz}), we can calculate the relationship between
the affine parameter and the redshift in this cosmology. It may be
written as follows,
\be
  \frac{dy}{dz}= \frac{1}{H_0 { \left( 1+z \right) }^{7/2}}.
  \lb{dydz}
\ee

\subsection{Volumes}\label{vols}

Before we proceed with the tests, there is one more issue to be taken
care of: volume definitions. This is a tricky issue in relativistic
cosmology as in general relativity volumes are not uniquely definable
quantities, and they may vary quite substantially from one another even
at $z \approx 0.1$ (Ribeiro 2001). Therefore, to avoid confusion, let us
state clearly what volume definitions we have been using.

Equation (\ref{Psi3}) states that both the luminosity function and
number density are defined using {\it local volumes}, that is, volume
elements defined {\it at the rest-frame} of a galaxy at a given redshift
$z$. {From} equations (\ref{dn3}) and (\ref{dl}) it is clear that
this volume is given by the following expression,
\be
  dV_{\rm local}=dl.dS_0=(1+z)dy { \left( d_{\sssty A} \right) }^2
  d \Omega_0.
  \lb{vlocal}
\ee
Considering equation (\ref{dydz}) it follows that in EdS cosmology
this reduces to,
\be
  dV_{\rm local}= { \left( d_{\sssty A} \right) }^2 d \Omega_0
  \frac{dz}{H_0 { (1+z) }^{5/2}}.
  \lb{vlocaleds}
\ee
Sometimes observers refer to their data being reduced against the {\it
real}, or proper volume, obtained by means of the spatial portion
of metric (\ref{eds}), as follows,
\be 
  dV_{\rm proper}=a^3r^2dr \sin \theta d \theta d \varphi.
  \lb{vreal}
\ee
Remembering equations (\ref{z2}) and (\ref{da3}), one can easily show
that proper and local volumes, that is, equations (\ref{vlocaleds})
and (\ref{vreal}), are {\it equal}. Nevertheless, we
believe that since both the luminosity function and number density are
defined locally, it makes more conceptual sense to refer to the local
volume, as given by equations (\ref{vlocal}) and (\ref{vlocaleds}). In
addition, it is important to stress that the local and proper volumes
are in general {\it not} equal to the observed volumes defined by
equation (\ref{v0}), as observed volumes use the three observational
distances given by the reciprocity theorem (\ref{recip}).

When carrying out their data reduction, observers also used to refer to a
{\it survey volume}, defined as (Sandage \etal 1979),
\be
   V_{\rm survey} = \frac{1}{3} d \Omega_0 { \left(
   \frac{{\cal U}_{\rm max}}{H_0} \right) }^3,
   \lb{vmax}
\ee
where ${\cal U}_{\rm max}$ is the maximum {\it velocity} obtained in the
survey, being calculated by means of the Doppler velocity-redshift
approximation ${\cal U}=cz$, valid only for small $z$. As it is
well known, it is only under this approximation that we can write the
redshift-distance law, that is, the Hubble law, as a velocity-distance
law (Harrison 1993), since the velocity-distance law is a general result
of an isotropic homogeneous expanding universe, whereas the
redshift-distance law is a mere approximation of the form, valid at
small redshift. That, in turn allows us to obtain the survey volume.
Such a procedure not only employs the low redshift approximation, but
also implicitly assumes the redshift distance $d_z$ as the distance
definition used in $V_{\rm survey}$ (see footnote at page \pageref{note5}).
Therefore, if one follows this path, model independent observable distances,
like $d_{\sssty A}$ or $d_{\sssty L}$, are related to $d_z$ only in an
approximate manner in EdS cosmological model, $d_z$ approximately
scales only with $d_{\sssty L}$, in addition to following the same
asymptotic behavior (Ribeiro 2001). 

Finally and most importantly, nowadays observers mostly use the comoving
volume when determining LF parameters of galaxy redshift survey data. In this
paper we have been using the proper volume in all calculations because
it is more conceptually natural. Thus, we require a conversion factor
to fit the CNOC2 data into our equations. Defining a comoving volume
requires a metric function in comoving coordinates. In EdS cosmology
it is obvious from equations (\ref{scalefactor}) and (\ref{vreal}) that,
\be
  dV_{\rm proper}= \frac{4}{9 {H_0}^2} { \left( 1+z \right) }^{-3}
  dV_{\rm comoving},
  \lb{conversion}
\ee
which shows clearly the conversion factor necessary in our calculations.

\subsection{Consistency Equations}

Before we write down the consistency equations for the LF parameters
with respect to the underlying cosmology they assume, let us
first briefly summarize the general formalism adopted so far. Let $T$
be some theoretical quantity completely determined by theory, and $T_0$
its measurement. Then, if $J$ is the completeness parameter, solely
written in terms of the LF, we have that
\be
   T_0=JT.
   \lb{t1}
\ee
The observational quantity is obtained at each bandwidth $W$. So,
\be
   T_0= \sum_{\sssty W} a_{\sssty W} T_0^{\sssty W},
   \lb{t2}
\ee
where $T_0^{\sssty W}$ is a quantity directly built from observations
($\psi$, for instance).

The dark matter is assumed to be in two components: (i) galactic dark
matter, accounted for by $J$, as it ought to be included when measuring
${\cal M}_g$, and (ii) intergalactic, possibly non-baryonic, dark matter,
accounted for by the term $T_d$. So, $T_d$ does not include any kind of
galactic dark matter. In addition, if there is intergalactic dark matter,
it must be added to $T_0$ in order make up for the whole energy-matter
as given by $T$. Consequently, equation (\ref{t1}) must be rewritten,
\be
   T= T_d + \sum_{\sssty W} a_{\sssty W} \frac{ T_0^{\sssty
      W}}{J^{\sssty W}}.
   \lb{t3}
\ee
% which means that we must have the following identity,
% \be
%    \frac{T_d}{J} \equiv T_d.
%    \lb{t4}
% \ee
Bearing those points in mind, we can now proceed in building our
consistency equations.

\subsection{First Consistency Equation: Local Density and Number
 Evolution}\lb{densisec} 

\subsubsection{Proper Volume}

Equations (\ref{n}), (\ref{Psi3}), and (\ref{rhoz}) will give the
theoretical behavior of the luminosity function in EdS cosmology.
In proper volume units, and in correct dimensions, it may be
written as,
\be
  \Psi {\cal M}_g= \frac{3 {H_0}^2}{8 \pi G} { \left( 1+z \right) }^3.
  \lb{Psi4}
\ee
Remembering equations (\ref{mg}), (\ref{Psi2}), (\ref{aw}), and
(\ref{bw}), and that the luminosity function is determined for each
galaxy morphological, or spectral, type, and at each observational
bandwidth, the observational counterpart of the equation above is
given by,
\be
  { \left( \Psi {\cal M}_g \right) }_0 = \sum_{{\sssty W}=1}^{\sssty
  W_f} a_{\sssty W} \left( \psi^{\sssty W} {\cal M}_g \right) = 
  \sum_{{\sssty W}=1}^{\sssty W_f} a_{\sssty W} \sum_{v=1}^{V}
  P^{\sssty W}_v(z) {\cal M}_v(z) \psi^{\sssty W}_v (z).
  \lb{Psi5}
\ee
These two equations provide the {\it first consistency test} between
Einstein-de Sitter theoretical prediction (eq.\ \ref{Psi4}) and
observations (eq.\ \ref{Psi5}) given by the LF. They basically
probe the behavior of the local density and number evolution of sources
at higher redshifts. Comparing them also gives us an estimate of
the intergalactic dark matter component. In fact, remembering equation
(\ref{t3}), its dependence on the redshift can be rewritten, 
\be
  { \rho }_d = \frac{3 {H_0}^2}{8 \pi G} { \left( 1+z
  \right) }^3 - \sum_{{\sssty W}=1}^{\sssty W_f} a_{\sssty W}
  \sum_{v=1}^{V} P^{\sssty W}_v(z) {\cal M}_v(z) \Psi^{\sssty W}_v (z).
  \lb{d1}
\ee
Notice that for other more general FLRW cosmologies, equation
(\ref{Psi4}) will have additional terms due to the deceleration parameter
$q_0$ and a possibly non-zero cosmological constant, while equation
(\ref{Psi5}) will remain the same, inasmuch as it is built from
observations.

\subsubsection{Comoving Volume}

For a comoving volume, we require the conversion factor (\ref{conversion})
to be applied to equation (\ref{Psi4}). In correct dimensions, it yields,
\be
 \Psi {\cal M}_g= \frac{{H_0}^2}{6 \pi G}.
 \lb{comove}
\ee
This is the quantity to be compared with the observational results as
given by the calculation of equation (\ref{Psi5}) from the CNOC2 data.

\subsection{Second Consistency Equation: Differential Number
            Counting}\label{46}

Considering equations (\ref{mz2}), (\ref{lm2}), (\ref{dydz}), and
remembering equation (\ref{t2}), we can write the observational
relativistic density per source, which is proportional to the
differential number counts, of an Einstein-de Sitter universe in the
volume units in which the LF is written, as follows,
\be
   {\mu}_0(z) = \frac{c}{H_0 { \left( 1+z \right) }^{5/2}}
   { \sum_{{\sssty W}=1}^{\sssty W_f} a_{\sssty W} \sum_{v=1}^{V}
   P^{\sssty W}_v(z) {\cal M}_v(z) \psi^{\sssty W}_v (z)}.
   \lb{mzeds}
\ee
This expression together with equation (\ref{meds}) provides the {\it
second consistency test} in EdS cosmology.

It is important to mention that equation (\ref{mzeds}) evaluates
${\mu}_0$ by means of the LF data, whereas if data were to be provided
on differential number counts, ${\mu}_0$ could  be calculated
by means of equation (\ref{mz0}), where fewer hypotheses are ``plugged''
in for the observational data. This is a point we want to stress. Equation
(\ref{mz0}) only gives differential number counts for either some
specific angular regions or over the whole sky, summed over all
wavebands, which means that there is {\it no} need to evaluate the proper
or comoving volume, as it is the case for the LF. Therefore, the best
prospect for observational cosmology tests which use redshift survey
data lies in the determination of the relativistic density per source
of the universe, which in turn requires differential number counting
observations as free as possible from any presupposed cosmological model.

As in \S \ref{densisec} above, the dark matter component associated
with the relativistic density per source of the universe may, in correct
dimensions, be written as,
\be
  {\mu}_d=\frac{3cH_0}{8 \pi G} \sqrt{1+z} - \frac{c}{H_0 { \left( 
  1+z \right) }^{5/2}}{ \sum_{{\sssty W}=1}^{\sssty W_f} a_{\sssty W}
  \sum_{v=1}^{V} P^{\sssty W}_v(z) {\cal M}_v(z) \Psi^{\sssty W}_v (z)}.
  \lb{meds2}
\ee

It is also worth mentioning that equations (\ref{avg}) and
(\ref{ml0}) provide ways of measuring the average density and the
mass-to-luminosity ratio, respectively. However, from a theoretical
viewpoint they are just extensions of the two tests above, with no
new quantities involved. Therefore, we shall leave their testing
to future work.

\subsubsection{Comoving Volume}

As with the local density, to obtain the theoretical relativistic
density per source
of the Universe in comoving volume units we need to apply the conversion
factor (\ref{conversion}) to its proper volume counterpart. From equations
(\ref{meds}) and (\ref{conversion}), it is clear that the theoretical
prediction of the model for $\mu$ (comoving) may be written as,
\be
  {\mu} (z)=\left[ \frac{3cH_0}{8 \pi G} \sqrt{1+z} \right] 
            \left[ \frac{4}{9 { \left( 1+z \right) }^3} \right]
           =\frac{c H_0}{6 \pi G} { \left( 1+z \right) }^{-5/2},
  \lb{dois}
\ee
whereas the observational quantity is just equation (\ref{mzeds}), since 
Lin \etal (1999) CNOC2 LF parameters are already written in comoving
volume units.

\section{Consistency Tests with the CNOC2 Data}\lb{testing}

Before we start to actually test the CNOC2 data against the predictions
of the EdS model, we still need to describe the means of evaluating equations
(\ref{psi}), (\ref{Psi}), and (\ref{j3}) in some waveband $W$. The
critical equation in this respect is the waveband limited version of
the selection function (eq.\ \ref{psi}), since the other two come from
this one. Its waveband version may be written as follows, 
\be
   \psi^{\sssty W} (z) =\int_{L_{\sssty W}(z)/{L_\ast}}^{\infty}
   \phi^{\sssty W} ( \ell ) d \ell.
   \lb{psiw}
\ee
Here $L_{\sssty W}(z)$ is the minimum absolute luminosity in the
filter waveband $W$, as determined at the observer, of sources detected
in the given sample. The source spectral function, ${\cal J} \left[ \nu
(1+z) \right]$, the proportion of radiation it emits in each frequency
at the source itself, is also observed here, but corrected by the factor
$(1+z)$ to account for its cosmological redshift. By examining equation
(\ref{fw}) one can only conclude that $L_{\sssty W}(z)$ must be written
as follows,
\be 
   L_{\sssty W}(z)= \int_0^\infty L(z) W(\nu) {\cal J} \left[ \nu (1+z)
    \right] d\nu.
    \lb{lw}
\ee
The luminosity evolution of sources is represented by their intrinsic
luminosity as a function of redshift, $L(z)$, as given in its rest frame,
multiplied by ${\cal J} \left[ \nu (1+z) \right]$. But, without some
theory for intrinsic luminosity evolution, the observer can only have
access to the product $ \left\{ L(z) {\cal J} \left[ \nu (1+z) \right]
\right\}$ as seen through the filter $W$. In principle, ${\cal J}$ is
also a function of the morphology and age of the source, and this is
reflected in samples which make distinctions on the basis of source
morphology and spectral type.

Equation (\ref{fw}) may be rewritten in terms of absolute magnitudes
$M_{\sssty W}(z)$ if we consider the expression above, and the
reciprocity theorem (\ref{recip}), yielding,
\be 
   L_{\sssty W}(z)= \frac{4 \pi { \left( 10 {\rm pc} \right) }^2}{(1+z)}
                    10^{0.4 \left[ C_{\sssty W} - M_{\sssty W}(z)
		    \right] },
   \lb{lw2}
\ee
where $C_{\sssty W}$ is the zero-point-magnitude-scale constant relative
to the filter $W$. This is a general equation, valid for any
cosmological model. The $(1+z)$ factor appearing in the denominator
accounts for the relativistic correction, as compared to the Euclidean case.
Schechter's LF (\ref{sch}) together with equation (\ref{lw2})
allows us to numerically compute the selection function by means of simple
quadratures of equation (\ref{psiw}).

Alternatively, the selection function can also be evaluated if we consider
equation (\ref{sch2}) for Schechter's LF written in terms of the absolute
magnitude. Thus, we can rewrite the $W$-filter selection function, as follows,
\be
   \psi^{\sssty W} (z) =\int^{M_{\sssty W}(z)}_{-\infty}
   \left( 0.4 \ln 10 \right) \phi_\ast 10^{0.4 \left( 1+\alpha
   \right) \left( M_\ast - {\overline{M}}_{\sssty W} \right) } \exp
   \left[ -10^{0.4 \left( M_\ast - {\overline{M}}_{\sssty W} \right)}
   \right] d \; {\overline{M}}_{\sssty W}. 
   \lb{psiw2}
\ee
Considering equation (\ref{evol}), which models the luminosity evolution
in the CNOC2 survey, this equation takes the form,
\begin{eqnarray}
   \psi^{\sssty W} (z) & = & 0.4 \ln 10 \; \phi_\ast \lb{psiw3} \\ 
   & \times &  \int^{M_{\sssty W}(z)}_{-\infty} 10^{0.4 \left( 1+ \alpha
   \right) \left[ M_\ast^\prime - Q \left( z-0.3 \right) -
   {\overline{M}}_{\sssty W} \right] } \exp \left\{  -10^{0.4 \left[
   M_\ast^\prime - Q \left( z-0.3 \right) - {\overline{M}}_{\sssty W}
   \right] } \right\} d \; {\overline{M}}_{\sssty W}. \nonumber
\end{eqnarray}
 
In either case $\phi_\ast$, $M_\ast^\prime$, $Q$, $\alpha$ will take the
fitted values as given in tables 1-3, the absolute magnitude range for
each filter used in the LF derivations of the CNOC2 sample, for {\it Hubble
constant} $h=1$, are, $-23.0 < M_{\sssty R_c} < -17.0$, $-22.0 < M_{\sssty U}
< -16.0$, $-22.0 < M_{\sssty AB} < -16.0$ (Lin \etal 1999, tables 1-2),
and the magnitude zero-point constant $C_{\sssty W}$ of equation (\ref{lw2})
in each filter are, $C_{\sssty R_c}=-13.64$, $C_{\sssty U}=-14.06$, $C_{\sssty
AB}=-48.60$ (Fukugita, Shimasaku and Ichikawa 1995).

Once $\psi$ is known by either method above (we used the second method,
with absolute magnitudes, as it requires fewer data manipulations), we can
calculate the summation term for all filters and all galaxy morphologies,
\be
   D(z) \equiv \mathop{\sum_{{\sssty W}=1}^3}_{\sssty (R_c, U, AB)} b_{\sssty W}
         \; \sum_{v=1}^3 P^{\sssty W}_v {\cal M}_v \psi^{\sssty W}_v(z). 
   \lb{sum}
\ee
Here $b_{\sssty R_c}=b_{\sssty U}=b_{\sssty AB}=1/3$ (see eq.\ \ref{bw}),
since the same CNOC2 sample was used at different bandwidth observational
windows for obtaining the LF parameters. The dynamical masses for each
morphological type are ${\cal M}_1=0.5 \times 10^{11} {\cal M}_\odot$,
${\cal M}_2=0.3 \times 10^{11} {\cal M}_\odot$, ${\cal M}_3=0.1 \times
10^{11} {\cal M}_\odot$ (Sparke and Gallagher 2000, p.\ 204, 264; see
also \S \ref{AGM} above). The results are summarized in tables 4-7.

\subsection{First Test}

The first consistency test compares the theoretical predictions of the
local density in comoving coordinates (eq.\ \ref{comove}),
and its observational equivalent. They are respectively given by,
\be
   \Psi {\cal M}_g = \frac{{H_0}^2}{6 \pi G}= 37.8 \times 10^9 % \; \; \mbox{G}
   {\cal M}_\odot \; {\mbox{Mpc}}^{-3}h^3,
   \lb{teste1}
\ee
and (see eq.\ \ref{Psi5}),
\be
  { \left( \Psi {\cal M}_g \right) }_0=D.
  \lb{teste1a}
\ee
Results of the D-term are found in table 7. One can clearly see that
the observational results are approximately constant around 0.5,
which gives the luminous matter up to $z=1$, as detected, and
extrapolated, by the CNOC2 galaxy redshift survey, as being about
1.3\% of the critical mass of EdS universe model. Since this quantity is
approximately constant, one can conclude that the LF determination in
this catalogue is, at least in this respect, consistent with the
theoretical predictions. In addition, this percentage of luminous
matter as compared to the total mass seems to fall in line with
some of the current expectations that the Universe is filled with
large quantities of dark mass, much of it non-baryonic, and not
distributed as the galaxies are.

One can also see a gradual rise in $D$ which might have some marginal
significance. As equation (\ref{teste1}) indicates, theoretically
$\Psi {\cal M}_g$ should be constant. But, observationally, from the
LF parameters it is not because Lin \etal (1999) have included some
number density evolution, which is reflected in $\Psi_0$ (that is clear
from the discussion in their paper). Thus, $\Psi_0$ is a function of
redshift, and increases slightly with redshift. However, since we are
holding ${\cal M}_g$ constant, their product increases with redshift. In
order to keep the product constant we would have to give ${\cal M}_g$ a
dependence on redshift in order to cancel the dependence of $\Psi_0$
on it.

\subsection{Second Test}

\begin{figure}
\centering
%%%%begin{apj does not understand LaTeX pictures}
%\input{plot.tex}
\plotone{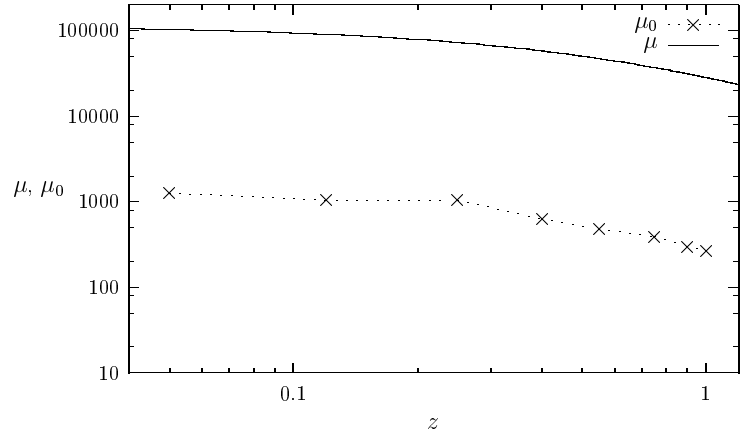}
\vspace{-13cm}
%%%%end{apj does not understand LaTeX pictures}
\caption{This plot shows the theoretical and observational results for
         the relativistic density of the universe (y-axis) vs.\ the
	 redshift. The lower curve, made of points joined by dots, shows
	 the observational results of $\mu_0$, derived from the CNOC2
	 redshift survey, whereas the full upper curve is the theoretical
	 prediction for $\mu(z)$ in EdS cosmology. One can clearly see
	 that although $\mu_0$ has the same downward tendency as $\mu$,
	 the theoretical results decrease more rapidly in the range $0.1
         \apl z \apl 0.4$. For $z>0.4$ an equivalent-to-the-theoretical-values
         downward trend is resumed. At $z=0.05$ $\mu_0$ is about 1.3\%
         of $\mu$, becoming 1.6\% at $z=0.25$, only to return to be 1.3\%
         of $\mu$ at $z=0.4-0.55$. Notice that the CNOC2 sample used in
         the LF parameters determination went only up to $z=0.55$, and
         the LF results were considered valid in the range $0<z<1$
         by means of extrapolations (see \S \ref{cnoc2} above, and Lin
         \etal 1999). These results indicate a broad qualitative agreement
         between theory and observations, apart from the region $0.1 \apl
         z \apl 0.4$, which may imply that either the
         chosen EdS cosmology provides an inadequate representation of
         the observational differential number counting data, or the
         CNOC2 survey detected an excess number of galaxies in most of
         its nominal observed redshift bin of $0.12 < z < 0.55$ as
         compared to EdS predictions.} \label{plot}
\end{figure}

The second test involves comparing the theoretical expression for the
relativistic density per source in a comoving volume, which is
equivalent to number counts, as given by equation (\ref{dois}), and
its observationally-derived-LF expression
(eq.\ \ref{mzeds}), which may also be written as,
\be
   \mu_0 (z)=\frac{cD}{H_0 { \left( 1+z \right) }^{5/2}}.
   \lb{teste2}
\ee
The functional relationship of the observational and theoretical results
against the redshift are respectively given in the third and fourth rows
of table 7, where one can clearly see a similar downward trend for both
quantities at higher redshifts. Nevertheless, the theoretical predictions
do tend to decrease more quickly than the observational results. At the
range $0.1 \apl z \apl 0.25$ the observational results tend to level off,
restarting a decrease only at $z \apg 0.3$ (see figure \ref{plot}).
Since $\mu_0$ gives information about the differential number counts
(see eqs.\ \ref{mz2} and \ref{mz0}), it seems thus reasonable to
conclude that the theory does not adequately account for the galaxy
counts at higher redshifts. This may reflect either the inadequacy of
the EdS model or of the number evolution model Lin \etal (1999) are
employing in the calculation of their LF's.

\section{Conclusions}\lb{concluding}

In this paper we have discussed the theory connecting the relativistic
cosmology number count theory with the astronomical data, practice and
theory behind the galaxy luminosity function (LF). We started from
Ellis' (1971) general, and model independent, relativistic expression
for number counts at a point along the observer's null cone and derived
expressions for the relativistic density per source of the universe,
differential number counts, mass-to-luminosity ratio, and local and
average densities. We then linked those expressions to the current
definition adopted in observational cosmology for the luminosity
and selection functions. The equations were constructed in such a way
as to reflect the fact that the luminosity function is determined
only within certain bandwidth ranges, and for some galaxy morphological
types.

We then discussed the current determination of the LF parameters 
in order to connect the theory to current practice.  The resulting
theoretical-observational framework for confronting theory with
cosmologically relevant observational results can be used in a number
of ways. We can factor the cosmology out of the LF's to obtain the
differential number counts; we can investigate the range of validity
of various low-redshift and Euclidean approximations; as we
have done here, we can take LF results and quickly check their
consistency with the cosmology they assume for a given range of 
redshift; we can use LF number counts results, with or without
assumed models of luminosity and number density evolution, and
either fit them to a given cosmology or calculate the cosmologies
they determine. In order to accomplish the latter we also need observer
area distance (or luminosity distance) vs.\ redshift data. As we
mentioned in passing, an analogous theoretical-observational 
framework can be easily constructed for this key cosmological
parameter. In this paper, as a simple example of how the general
theoretical-observational framework can be used, we examined the
consistency between the LF parameters of a redshift survey and
some of the key equations of the cosmology they assume. We chose the
CNOC2 intermediate redshift survey, and the EdS determined parameters
that Lin \etal (1999) calculated from it, to illustrate this in
a simple way. After specializing our general model to EdS, we carried
out two consistency tests, one related to the local density and number
evolution, and another dealing with the differential number counting,
comparing the theoretical predictions with the observational results
derived from the LF determination of the CNOC2 survey. The results
show a general agreement between theory and observation in both
tests. Nevertheless, the second test indicated a possible excess of
galaxies at $0.1 \le z \le 0.4$, which either means that the EdS
cosmology does not accurately represent the data, or that the
evolution model assumed by Lin \etal (1999) is not adequate.

It is important to point out that the actual observations used in the
CNOC2 survey LF determination were carried out only within the range
$0.12 < z < 0.55$, which means that in most of its observational
redshift interval there appears to be an excess of galaxies as
compared to EdS predictions. For $z>0.4$ our results indicate that
this excess tends to disappear. In order to account for this extra
mass we can either assume that the universe has a closed FLRW
geometry, or hypothesize, as many others have been doing recently,
a non-zero cosmological constant as a possible extra source of
mass-energy, or modify the model of number evolution Lin \etal (1999)
are employing.

The fourth option, a possible inadequacy of the EdS cosmological model,
appears to us less likely, since the LF determinations seem to have
only a weak dependence on the chosen model. If we were to be suspicious
about the inadequacy of the cosmological model, due to this weak
dependency we should perhaps call into question not the EdS universe,
but the whole set of standard FLRW cosmologies. It seems to us that our
results do not support such a more radical viewpoint, but this is
something which one always need to bear in mind.

%%%%%%%%%%%%%%%%%%%%%%%%%%%%%%%%%%%%%%%%%%%%%%%%% acknowledgements
\acknowledgments
We are grateful to Chris Impey for illuminating discussions on the
methodologies behind data reduction and analysis of redshift surveys,
to Huan Lin for clarifications regarding the CNOC2 survey, and a referee
for useful remarks. MBR wishes to thank the Vatican Observatory Research
Group for their kind hospitality while performing this research, and to
acknowledge the financial support from Brazil's CAPES Foundation. 
%%%%%%%%%%%%%%%%%%%%%%%%%%%%%%%%%%%%%%%%%%%%%%%%% References
%% cross-referencing. Each reference is preceded by a \bibitem command
%% that defines in curly braces the KEY that corresponds to the KEY in
%% the \cite commands. Make sure that you provide a unique KEY for
%% every \bibitem or else the paper will not LaTeX. The square brackets
%% should contain the citation text that LaTeX will insert in place of
%% the \cite commands.

\clearpage
\clearpage 
\clearpage 
% table 1
\begin{deluxetable}{cccccc}
\tabletypesize{\normalsize}
\tablecaption{LF fit in CNOC2 sample $B_{\sssty AB}$ band
             ($q_0=0.5$). \lb{table1}}
\tablewidth{0pt}
\tablehead{
\colhead{$v$} & \colhead{$\alpha$} & \colhead{$\phi_\ast$} &
\colhead{$Q$} & \colhead{$M_\ast^\prime$} & \colhead{$P_v$}
}
\startdata
   1  &   0.08   &   0.0203    & 1.58 &       -19.06   & 0.29 \\ 
   2  &   -0.53  &   0.0090    & 0.90 &       -19.38   & 0.24 \\ 
   3  &   -1.23  &   0.0072    & 0.18 &       -19.26   & 0.47 \\ 
\enddata
\end{deluxetable}
\clearpage 
% table 2
\begin{deluxetable}{cccccc}
\tabletypesize{\normalsize}
\tablecaption{LF fit in CNOC2 sample $R_c$ band ($q_0=0.5$). \lb{table2}}
\tablewidth{0pt}
\tablehead{
\colhead{$v$} & \colhead{$\alpha$} & \colhead{$\phi_\ast$} &
\colhead{$Q$} & \colhead{$M_\ast^\prime$} & \colhead{$P_v$}
}
\startdata
   1  &   -0.07  &   0.0185    & 1.24 &       -20.50   & 0.29 \\ 
   2  &   -0.61  &   0.0080    & 0.69 &       -20.47   & 0.24 \\ 
   3  &   -1.34  &   0.0056    & 0.11 &       -20.11   & 0.47 \\ 
\enddata
\end{deluxetable}
\clearpage 
% table 3
\begin{deluxetable}{cccccc}
\tabletypesize{\normalsize}
\tablecaption{LF fit in CNOC2 sample $U$ band ($q_0=0.5$). \lb{table3}}
\tablewidth{0pt}
\tablehead{
\colhead{$v$} & \colhead{$\alpha$} & \colhead{$\phi_\ast$} &
\colhead{$Q$} & \colhead{$M_\ast^\prime$} & \colhead{$P_v$}
}
\startdata
   1  &   0.14   &   0.0213    & 1.85 &       -18.54   & 0.29 \\ 
   2  &   -0.51  &   0.0092    & 0.97 &       -19.27   & 0.24 \\ 
   3  &   -1.14  &   0.0095    & 0.51 &       -19.32   & 0.47 \\ 
\enddata
\end{deluxetable}
\clearpage 
% table 4
\begin{deluxetable}{cccc}
\tabletypesize{\normalsize}
\tablecaption{CNOC2 selection function\tablenotemark{a} \ results in
              the $R_c$ band  vs.\ redshift.\tablenotemark{b}
	      \lb{table4}}
\tablewidth{0pt}
\tablehead{
\colhead{$z$} & \colhead{$\psi_1^{\sssty {R_c}}$} &
\colhead{$\psi_2^{\sssty {R_c}}$} & \colhead{$\psi_3^{\sssty {R_c}}$}
}
\startdata
  0.05 & 0.0181 & 0.0120 & 0.0220 \\ 
  0.12 & 0.0182 & 0.0121 & 0.0221 \\ 
  0.25 & 0.0183 & 0.0123 & 0.0223 \\ 
  0.4  & 0.0185 & 0.0125 & 0.0225 \\ 
  0.55 & 0.0186 & 0.0127 & 0.0227 \\ 
  0.75 & 0.0187 & 0.0130 & 0.0230 \\ 
  0.9  & 0.0187 & 0.0132 & 0.0232 \\ 
  1.0  & 0.0187 & 0.0133 & 0.0234 \\ 
\enddata
\tablenotetext{a}{Subscript numbers denote the spectral type
                  morphology adopted for the CNOC2 survey (see \S
		  \ref{morphology}). Units are Mpc$^{-3}$.} % h$^3$.}
\tablenotetext{b}{Redshift values were taken to agree with the
                  nominal redshift interval and bins of the survey,
		  and the claimed validity of the fitted LF parameters
		  (see \S \ref{cnoc2} above, and Lin \etal 1999, p.\
		  550).}
\end{deluxetable}
\clearpage 
% table 5
\begin{deluxetable}{cccc}
\tabletypesize{\normalsize}
\tablecaption{CNOC2 selection function results in the $U$ band vs.\
             redshift.\lb{table5}}
\tablewidth{0pt}
\tablehead{
\colhead{$z$} & \colhead{$\psi_1^{\sssty U}$} &
\colhead{$\psi_2^{\sssty U}$} & \colhead{$\psi_3^{\sssty U}$} 
}
\startdata
  0.05 & 0.0180 & 0.0119 & 0.0284 \\ 
  0.12 & 0.0182 & 0.0121 & 0.0289 \\ 
  0.25 & 0.0186 & 0.0123 & 0.0297 \\ 
  0.4  & 0.0189 & 0.0126 & 0.0307 \\ 
  0.55 & 0.0192 & 0.0128 & 0.0317 \\ 
  0.75 & 0.0194 & 0.0131 & 0.0330 \\ 
  0.9  & 0.0195 & 0.0134 & 0.0341 \\ 
  1.0  & 0.0196 & 0.0135 & 0.0348 \\ 
\enddata
\end{deluxetable}
\clearpage 
% table 6
\begin{deluxetable}{cccc}
\tabletypesize{\normalsize}
\tablecaption{CNOC2 selection function results in the $B_{\sssty AB}$
              band vs.\ redshift.\lb{table6}}
\tablewidth{0pt}
\tablehead{
\colhead{$z$} & \colhead{$\psi_1^{\sssty {AB}}$} &
\colhead{$\psi_2^{\sssty {AB}}$} & \colhead{$\psi_3^{\sssty {AB}}$}
}
\startdata
  0.05 & 0.0182 & 0.0122 & 0.0252 \\ 
  0.12 & 0.0183 & 0.0123 & 0.0254 \\ 
  0.25 & 0.0185 & 0.0125 & 0.0257 \\ 
  0.4  & 0.0187 & 0.0128 & 0.0260 \\ 
  0.55 & 0.0189 & 0.0130 & 0.0264 \\ 
  0.75 & 0.0190 & 0.0133 & 0.0268 \\ 
  0.9  & 0.0191 & 0.0135 & 0.0272 \\ 
  1.0  & 0.0191 & 0.0136 & 0.0274 \\ 
\enddata
\end{deluxetable}
\clearpage 
% table 7
\begin{deluxetable}{cccc}
\tabletypesize{\normalsize}
\tablecaption{D-term\tablenotemark{a} \ sum (eq.\ \ref{sum}) results
              vs.\ redshift.\lb{table7}}
\tablewidth{0pt}
\tablehead{
\colhead{$z$} & \colhead{$D$} &
\colhead{${ \left( 1+z \right) }^{-5/2}D$} & \colhead{${ \left(
1+z \right)}^{-5/2}$~\tablenotemark{b}}
}
\startdata
  0.05 & 0.47 & 0.42 & 0.89 \\ 
  0.12 & 0.47 & 0.35 & 0.75 \\ 
  0.25 & 0.48 & 0.35 & 0.72 \\ 
  0.4  & 0.49 & 0.21 & 0.43 \\ 
  0.55 & 0.49 & 0.16 & 0.33 \\ 
  0.75 & 0.50 & 0.13 & 0.25 \\ 
  0.9  & 0.51 & 0.10 & 0.20 \\ 
  1.0  & 0.51 & 0.09 & 0.18 \\ 
\enddata
\tablenotetext{a}{Units are $10^9{\cal M}_\odot$ Mpc$^{-3}$.} % h$^3$.}
\tablenotetext{b}{Included for comparison (see eq.\ \ref{dois}).} 
\end{deluxetable}
\end{document}